\documentclass[twocolumn]{emulateapj}
\usepackage{psfig, lscape}

\submitted{Version: \today ~~ 
Accepted: 2009 03 19}

\def\alwaysmath#1{\ifmmode{#1}\else{$#1$}\fi}
 
\shortauthors{Men\'{e}ndez-Delmestre et al.}
\shorttitle{Mid-IR Spectroscopy of Submillimeter Galaxies}

\begin{document}

\title{Mid-Infrared Spectroscopy of Submillimeter Galaxies:\\ Extended Star Formation in Massive High Redshift Galaxies} \author{\sc Kar\'{i}n
Men\'{e}ndez-Delmestre,\altaffilmark{1,2} Andrew W.\ Blain,\altaffilmark{2} Ian Smail,\altaffilmark{3} Dave M.\ Alexander\altaffilmark{4},  Scott C.\ Chapman,\altaffilmark{5} 
Lee Armus,\altaffilmark{6} Dave Frayer,\altaffilmark{6} Rob J.\ Ivison\altaffilmark{7,8} \& Harry Teplitz\altaffilmark{6}}

\altaffiltext{1}{NSF Astronomy and Astrophysics Postdoctoral Fellow}
\altaffiltext{2}{California Institute of Technology, MC 105-24, Pasadena, CA 91125}

\altaffiltext{3}{Institute for Computational Cosmology, Durham University, Durham DH1 3LE, UK}

\altaffiltext{4}{Department of Physics, Durham University, Durham DH1 3LE, UK}

\altaffiltext{5}{Institute of Astronomy, Madingley Road, Cambridge, CB3 0HA, UK}

\altaffiltext{6}{Spitzer Science Center, MC 220-6, California Institute of Technology, Pasadena, CA 91125}

\altaffiltext{7}{UK Astronomy Technology Centre, Blackford Hill, Edinburgh EH9 3HJ, UK}

\altaffiltext{8}{Institute for Astronomy, Blackford Hill, Edinburgh EH9 3HJ, UK}

\email{km@astro.caltech.edu}

\begin{abstract}
We used the {\it Spitzer} Infrared Spectrograph (IRS) to study the mid-IR properties and investigate the energetics of 24 Submillimeter Galaxies (SMGs). This is the largest sample of SMGs observed with {\it Spitzer} IRS and covers the full extent of properties of the radio-identified population of SMGs in the redshift range of $z \sim 0.65$--3.2. We observe broad emission features from Polycyclic Aromatic Hydrocarbons (PAHs) in more than 80\% of our sample.  We find that the median mid-IR spectrum is well described by a starburst component with an additional power law, $F_{\nu} \sim \nu^{-2}$, likely representing a $< 32$\% AGN contribution to the bolometric luminosity.  Our results thus confirm that starburst activity dominates the bolometric luminosity in SMGs. We find that SMGs show weaker silicate absorption at $\sim 9.7 \mu$m than local ULIRGs. We also find stronger 6.2-$\mu$m PAH emission in SMGs (relative to the 7.7$\mu$m PAH feature) than in local nuclear starbursts, which may be attributed to lower extinction by ice along the line of sight to SMGs. This suggests that the continuum and PAH emitting regions of SMGs are less obscured than in local starbursts and similarly luminous low-redshift ULIRGs.  We interpret these results as evidence for a more extended distribution of cool and warm dust in SMGs compared to the more compact emitting regions in local ULIRGs and starbursts. Together these results suggest that SMGs are not simple high-redshift analogs of local ULIRGs or nuclear starbursts, but instead they appear to have star formation which resembles that seen in less-extreme star-forming environments at $z\sim 0$ -- suggesting their intense activity is distributed across a far larger region than the $\sim 1$-kpc nuclear bursts in local ULIRGs.
\end{abstract}


%
%

\section{Introduction}

It is over a decade since the discovery of a population of high-redshift galaxies identified through their submillimeter (submm) emission \citep{smail97, barger99, eales99, bertoldi00, cowie02, scott02, borys03, webb03, coppin05, younger07}.  Their submm selection suggests that these are strongly star-bursting systems. With implied star-formation rates (SFRs) of $\sim 100$--1000\,M$_{\odot}$\,yr$^{-1}$, these dust-enshrouded {\it submm} galaxies (hereafter, SMGs) may make a significant contribution to the global SFR density at $z \sim 2$--3 (\citealt{chapman05}, hereafter C05). The absorbing dust that makes them such prodigious submm emitters also makes them quite optically faint and renders  follow-up studies at shorter wavelengths challenging. The study of SMGs has been facilitated by the detection of a large fraction of them as $\mu$Jy radio sources \citep{ivison02} and more recently through millimeter-wave interferometry. This has allowed the subsequent measurement of their spectroscopic redshifts for fairly large samples (C05).

%
%
\begin{deluxetable*}{lcccccc}
\tabletypesize{\scriptsize}
\tablecaption{Summary of {\it Spitzer} IRS Observations} 
\tablehead{
  \colhead{Name}  &  \colhead{R.A.$^a$} &  \colhead{Dec.} &  \colhead{$z_{opt}^b$} & \colhead{Date} &  \colhead{LL1} &  \colhead{LL2}    \\
\colhead{}  &  \colhead{($h\,m\,s$)} &  \colhead{($\circ\,'\,''$)} & \colhead{} & \colhead{}  &  \colhead{(ramp~[s]~$\times$~cycles)} &  
\colhead{(ramp~[s]~$\times$~cycles)}
\label{Obstab}}
\startdata
SMM\,J030227.73 & 03\,02\,27.73 & +00\,06\,53.5 & 1.408 & Jan.\ 2006 & 2 $\times$ 120 $\times$ 30 & 2 $\times$ 120 $\times$ 30 \\
SMM\,J030231.81 & 03\,02\,31.81 & +00\,10\,31.3 & 1.316 & Feb. 2006 & 2 $\times$ 120 $\times$ 30 & 2 $\times$ 120 $\times$ 30 \\
SMM\,J105151.69 & 10\,51\,51.69 & +57\,26\,36.0 & 1.147 & May 2006 & 2 $\times$ 120 $\times$ 30 & 2 $\times$ 120 $\times$ 30 \\
SMM\,J105158.02 & 10\,51\,58.02 & +57\,18\,00.3 & 2.239 & May 2006 & 2 $\times$ 120 $\times$ 30 & -- \\
SMM\,J105200.22 & 10\,52\,00.22 & +57\,24\,20.2 & 0.689 & May 2006 & 2 $\times$ 120 $\times$ 30 & 2 $\times$ 120 $\times$ 30 \\
SMM\,J105227.58 & 10\,52\,27.58 & +57\,25\,12.4 & 2.142 & May 2006 & 2 $\times$ 120 $\times$ 30 & -- \\
SMM\,J105238.19 & 10\,52\,38.19 & +57\,16\,51.1 & 1.852 & May 2006 & 2 $\times$ 120 $\times$ 30 & 120 $\times$ 30 \\
SMM\,J105238.19 & 10\,52\,38.30 & +57\,24\,35.8 & 3.036 & May 2006 & 2 $\times$ 120 $\times$ 30 & -- \\
SMM\,J123549.44 & 12\,35\,49.44 & +62\,15\,36.8 & 2.203 & May 2006 & 2 $\times$ 120 $\times$ 30 & -- \\
SMM\,J123553.26 & 12\,35\,53.26 & +62\,13\,37.7 & 2.098 & May 2006 & 2 $\times$ 120 $\times$ 30 & -- \\
SMM\,J123707.21 & 12\,37\,07.21 & +62\,14\,08.1 & 2.484 & May 2006 & 2 $\times$ 120 $\times$ 30 & -- \\
SMM\,J123711.98 & 12\,37\,11.98 & +62\,13\,25.7 & 1.992 & May 2006 & 2 $\times$ 120 $\times$ 30 & 2 $\times$ 120 $\times$ 30 \\
SMM\,J123721.87 & 12\,37\,21.87 & +62\,10\,35.3 & 0.979 & May 2006 & 2 $\times$ 120 $\times$ 30 & 2 $\times$ 120 $\times$ 30 \\
SMM\,J163639.01 & 16\,36\,39.01 & +40\,56\,35.9 & 1.495 & Aug.\ 2005 & 2 $\times$ 120 $\times$ 30 & 2 $\times$ 120 $\times$ 30 \\
SMM\,J163650.43 & 16\,36\,50.43 & +40\,57\,34.5 & 2.378 & Sept.\ 2005 & 2 $\times$ 120 $\times$ 30 & -- \\
SMM\,J163658.78 & 16\,36\,58.78 & +40\,57\,28.1 & 1.190 & Sept.\ 2005 & 2 $\times$ 120 $\times$ 30 & 2 $\times$ 120 $\times$ 30 \\
SMM\,J221733.02 & 22\,17\,33.02 & +00\,09\,06.0 & 0.926 & June 2006 & 2 $\times$ 120 $\times$ 20 & 2 $\times$ 120 $\times$ 30 \\
SMM\,J221733.12 & 22\,17\,33.12 & +00\,11\,20.2 & 0.652 & Nov.\ 2006 & 2 $\times$ 120 $\times$ 15 & 2 $\times$ 120 $\times$ 15 \\
SMM\,J221733.91 & 22\,17\,33.91 & +00\,13\,52.1 & 2.555 & June 2006 & 2 $\times$ 120 $\times$ 30 & -- \\
SMM\,J221735.15 & 22\,17\,35.15 & +00\,15\,37.2 & 3.098 & June 2006 & 2 $\times$ 120 $\times$ 30 & -- \\
SMM\,J221735.84 & 22\,17\,35.84 & +00\,15\,58.9 & 3.089 & June 2006 & 2 $\times$ 120 $\times$ 30 & -- \\
SMM\,J221737.39 & 22\,17\,37.39 & +00\,10\,25.1 & 2.614 & June 2006 & 2 $\times$ 120 $\times$ 30 & -- \\
SMM\,J221804.42 & 22\,18\,04.42 & +00\,21\,54.4 & 2.517 & June 2006 & 2 $\times$ 120 $\times$ 30 & -- \\
SMM\,J221806.77 & 22\,18\,06.77 & +00\,12\,45.7 & 3.623 & June 2006 & 2 $\times$ 120 $\times$ 30 & --  \\
\enddata
\tablenotetext{a}{radio position from C05 compilation.}
\tablenotetext{b}{spectroscopic redshifts from C05.}
\end{deluxetable*}

With a mean redshift of $<\! z\! > \sim 2.2$, the redshift distribution of the {\it radio-identified} SMGs in C05 coincides with the global peak epoch of quasar activity. The connection between SMGs and active galactic nuclei (AGNs) has been probed in recent near-IR and X-ray studies. Near-IR spectroscopy by \citet{swinbank04} show that broad H$\alpha$ lines (FWHM $\gtrsim 1000$ km\,s$^{-1}$) are often present in these galaxies, while deep X-ray studies using the sensitive {\it Chandra} Deep Field-North (CDF-N) Survey suggest that at least $\sim28$--50\% of SMGs host an obscured AGN \citep{alexander05}.  A number of SMGs that display no AGN signatures in the rest-frame optical \citep{swinbank04} were classified as AGNs based on X-ray observations \citep{alexander05}. This is likely due to geometrical effects in which the broad-line region of the AGN remains hidden in the  optical by intervening obscuring material.  At high X-ray energies, direct X-ray emission is detectable through even very high column densities, allowing the direct detection of AGN at very high obscurations. However, \citet{alexander05} find that the majority of AGNs in SMGs are more modestly obscured, with column densities of N$_H \gtrsim 10^{23}$\,cm$^{-2}$. A similar mix of AGN and starburst activity is seen in local  ultra-luminous infrared galaxies (ULIRGs, with total IR luminosities L$_{\rm{IR}} \gtrsim 10^{12}$ L$_\odot$; \citealt{soifer87, sanders88}), most of which have been shown to be composite AGN--starburst systems (e.g., \citealt{armus07}).  The SMGs have IR luminosities which are comparable to ULIRGs at low redshift, prompting the question as to whether SMGs are high-redshift analogs of ULIRGs and hence whether we can learn about the physical processes within SMGs from studies of local ULIRGs (\citealt{tacconi08}). 

The potential presence of luminous Compton-thick AGNs (N$_H \gtrsim 10^{24}$\,cm$^{-2}$) in those SMGs with no X-ray AGN signature remains a significant caveat to these results. Furthermore, the samples of SMGs with the necessary ultra-deep X-ray observations to reveal the presence of highly obscured  AGN are still small.  Hence it is possible that a small fraction of luminous, but Compton thick, AGN lurk within the SMG population (see \citealt{coppin08}).  

Rest-frame optical emission provides direct insight into the stellar emission and ionized gas of a galaxy, but suffers from obscuration due to intervening dust. At longer wavelengths, the mid-IR emission arising from the dust itself provides an indirect insight into the dust-enshrouded nature of SMGs and suffers from much less obscuration than the shorter, optical wavelengths. We therefore designed a program to follow up a large sample of 24 SMGs in the mid-IR using {\it Spitzer}'s Infrared Spectrograph (IRS; \citealt{houck04}) in an effort to answer the following questions for a large and representative sample of the SMG population:  Are SMGs composite AGN--starburst systems? To what extent does AGN activity contribute to the total infrared output of these galaxies? Is there a spectrum of varying levels of AGN activity across the population?  Are local ULIRGs good analogs for the mid-IR emission of SMGs?

%
%
\begin{figure*}
\plottwo{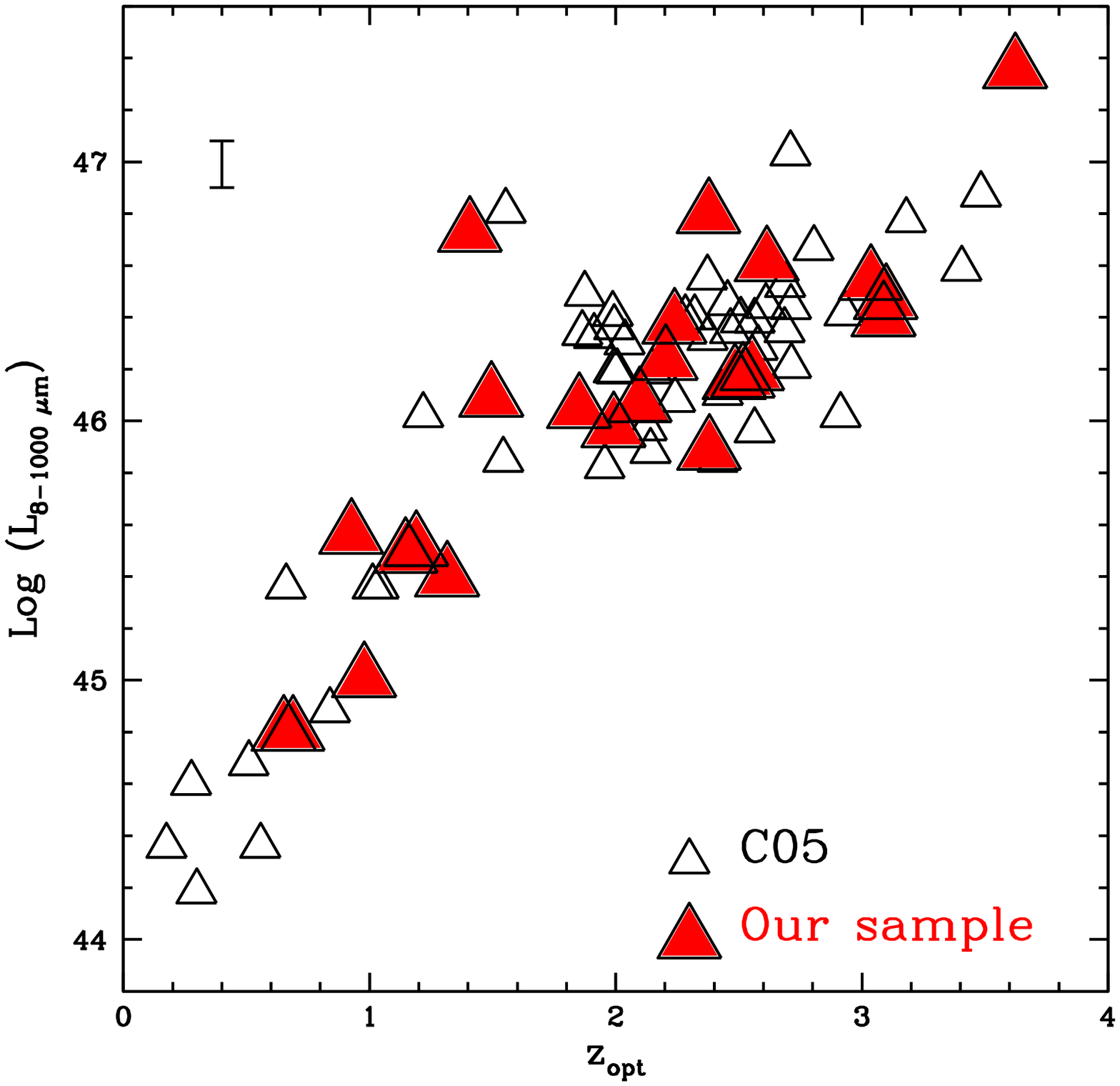}{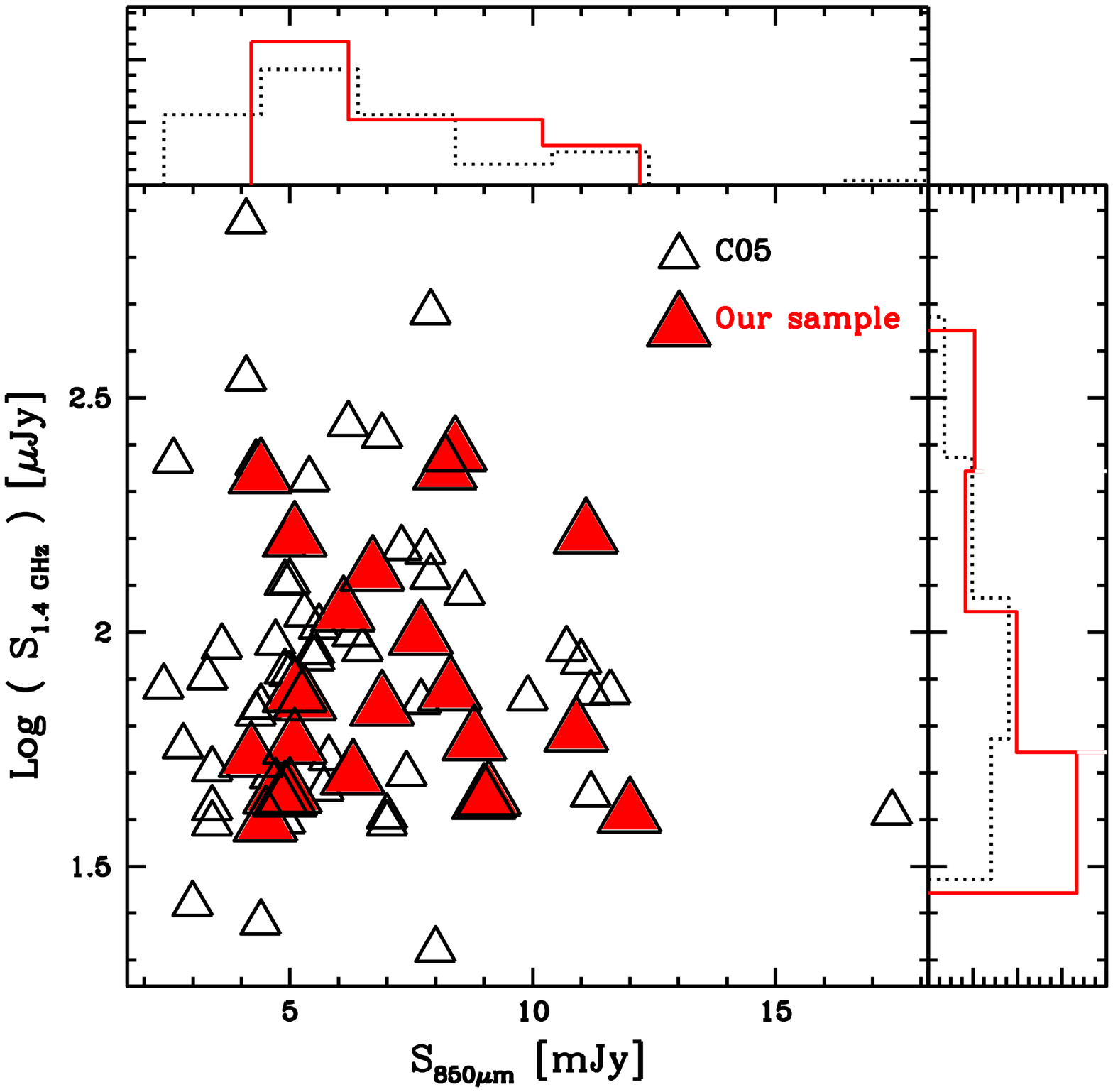}
\caption{({Left:}) The distribution of optical redshifts and IR luminosities (8--1000$\mu$m) of the parent sample of radio-identified SMGs compiled by C05 with our IRS subsample of SMGs highlighted.  We use IR luminosities based on the values reported by C05, corrected down by a factor of two, following \citet{kovacs06}. We plot a representative error on the IR luminosities, corresponding to $\sim 20\%$ (C05). ({Right:}) The distribution of radio and submillimeter fluxes for the full C05 sample and those in our subsample observed with IRS. The histograms show the distribution in parameters of the full sample (dotted line) and our IRS subsample (solid line). These plots demonstrate that the IRS sample is representative of the parent sample of radio-identified SMGs, covering the full range in radio and submm properties, redshift and IR-luminosity. 
\label{C05}}
\end{figure*}

The main components contributing to the mid-IR spectrum of a galaxy are: thermal dust continuum, emission from vibrational modes in Polycyclic Aromatic Hydrocarbons (hereafter, PAHs) and other atomic and molecular lines.  The continuum emission at longer mid-IR wavelengths, $\lambda \gtrsim 12\mu$m,  arises from emission by very small grains of dust (VSG; $\lesssim 10$nm) found around obscured AGN or star-forming regions. This is often referred to as the VSG continuum, or the {\it warm} dust continuum (T$_{Dust}\lesssim 250$\,K). At shorter wavelengths, $\lambda \lesssim 6\mu$m, the continuum traces emission from dust heated to significantly hotter temperatures (T$_{Dust} \gtrsim 500$\,K) likely due to its close location near an AGN or possibly a hot, nuclear starburst region. This is what we refer to as the {\it hot} dust continuum. PAH molecules ($\lesssim$ few nm) consist of chained benzene rings, associated hydrogen and other trace elements, such as Si and Mg. The line emission in the mid-IR waveband is dominated by PAH molecules which are excited by the UV photons that are copiously available in star-forming regions. The main PAH emission features arise from the bending and stretching of skeletal C-C or peripheral C-H bonds  and are observed at rest-frame 6.2, 7.7, 8.6, 11.3, 12.7 and 17 $\mu$m (e.g., \citealt{draine07}).  It has been shown locally that stronger PAH features are associated with regions of intense star formation  \citep{helou99}.

With the unprecedented sensitivity of {\it Spitzer} IRS, it has been possible to explore the mid-IR region of galaxies at high redshift down to continuum levels of $S_{24 \mu m} \gtrsim 0.1$ mJy (e.g., \citealt{yan05, yan07, lutz05, lutz07, desai06, rigopoulou06, teplitz07, siana08}). At lower redshifts, investigation of the mid-IR properties of local galaxies with IRS provides for a sample of detailed templates against which high-redshift sources can be compared (e.g., \citealt{spoon04, armus04, armus06, armus07, weedman05, brandl06, sturm06, desai07}).  A wide range of mid-IR spectra has been uncovered for high-redshift sources, ranging from continuum-dominated spectra with no PAH features to PAH-dominated spectra (e.g., \citealt{weedman08}).  Between these two extremes there is a myriad of composite spectra displaying a significant  continuum with superposed PAH features  (e.g., \citealt{yan07}). 

We presented the {\it Spitzer} IRS results for the first five SMGs observed in the program in \citet{md07}, which complemented earlier IRS results for two SMGs at $z \sim 2.8$ presented by \citet{lutz05}. Since then, \citet{valiante07} and \citet{pope08}, obtained IRS spectra of nine SMGs in blank field and cluster lens surveys and of 13 SMGs from the GOODS-North Field, respectively.  In this paper we present the results for our full observing program of 24 SMGs, comprising the largest sample of SMGs observed to date with {\it Spitzer} and nearly doubling the number of SMGs observed with IRS. We discuss our sample selection and observing strategy in \S\ref{obsred}. The steps comprising the reduction and analysis of the spectra are discussed in \S\ref{obsred} and \S\ref{analysis}.  Our results are presented in \S\ref{results} and discussed in detail in \S\ref{discuss}. We give our conclusions in \S\ref{conc}. We assume a $\Lambda$CDM cosmology, with $H_0 = 71$\,km\,s$^{-1}$\,Mpc$^{-1}$, $\Omega_M = 0.27$ and $\Omega_\Lambda = 0.73$.

%
%

\section{Observations and Reduction}\label{obsred}

The SMGs in our program were selected from the sample of 73 spectroscopically-confirmed, radio-identified SMGs in C05. We aimed for full coverage of the range in redshifts, radio, submm and total IR luminosities present in the C05 sample (see Fig.~\ref{C05}) to build a sample representative of the radio-identified SMG population in general.  The SMGs in our sample have optical redshifts $0.6 \lesssim z \lesssim 3.6$, radio and submm fluxes $40 \mu$Jy$\lesssim S_{1.4\rm{GHz}} \lesssim 240 \mu$Jy and  $4$ mJy$ \lesssim S_{850\mu\rm{m}} \lesssim 12$ mJy (C05), and IR luminosities $2 \times 10^{11}$L$_\odot \lesssim L_{\rm{8-1000}\mu\rm{m}} \lesssim 6 \times 10^{13}$L$_\odot$, where the latter are based on the bolometric values reported by C05, corrected down by a factor of two, following \citet{kovacs06}. We note that several of the SMGs from the C05 spectroscopic sample fall in the GOODS-North region and were included in the IRS studies focused on that field (\citealt{pope08}). Together with our sample, these are effectively part of the larger IRS study of the C05 survey which covers a total of 33 SMGs (or 45\% of the parent C05 sample).

While we aimed to span the full range in multiwavelength properties of radio-identifed SMGs, to have a realistic chance of detecting the galaxies with IRS we selected the 24 SMGs with higher estimated 24$\mu$m fluxes.\footnote{We used the online tool SPEC-PET to calculate {\it S/N} estimates (http://ssc.spitzer.caltech.edu/tools/specpet/).}  As {\it Spitzer} 24$\mu$m observations were not available at the time of the proposal, we estimated $24 \mu$m fluxes for the SMGs in C05 from fitting the spectral energy distributions (SEDs) to the radio (1.4\,GHz), submm ($850 \mu$m) and available optical photometric data points (see C05 and references therein). We assumed a power-law fit, $S_\nu \sim \nu^{-\alpha}$, to the mid-IR with $\alpha=-1.7$. We note that selection of $24 \mu$m-bright sources (albeit estimated) may introduce a two-sided selection effect.  Selection based on bright $24 \mu$m-fluxes will preferentially  pick out the hot dust emission in AGNs.  However, at redshifts $z \sim 2$, the $7.7 \mu$m-PAH feature is redshifted onto the $24 \mu$m-band: this is a prominent  emission feature, associated with star formation activity and will result in a boosted $24 \mu$m flux. We discuss this potential selection bias in more detail in \S\ref{RadioAND24umcomp}.  From a parallel mid-IR imaging survey of the radio-identified SMGs in C05 with the Multiband Imager for {\it Spitzer} (MIPS) by \citet{hainline09}  our IRS subsample has  MIPS 24~$\mu$m band fluxes in the range $S_{24~\mu\rm{m}} \sim 0.09$--0.85\,mJy with a median value of $<\! S_{24~\mu\rm{m}}\! > = 0.33 \pm 0.18$ mJy.

The details of our observations are summarized in Table~\ref{Obstab}. We observed all our targets using the low-resolution ($R \sim 57$--126) Long-Low (LL) observing module of IRS. The slit width for this module is $\sim 10.7 \arcsec$, which corresponds to roughly 90\,kpc at $z\sim2$, and so each target is fully enclosed in the observed slit and is thus treated as a point source. With shared coverage of the first LL order (LL1: 19.5--38.0$\mu $m) and the second LL order (LL2: 14.0--21.3$\mu$m) we aimed to cover rest-frame emission longwards of 6\,$\mu$m to search for PAH emission and for any silicate absorption at 9.7\,$\mu$m.  With a redshift range of $z=0.65$--3.2 for the full sample, the LL1 IRS module alone provides the relevant coverage for SMGs with $z \gtrsim 2.1$; while the LL2 module is also required for SMGs with $z \lesssim 2.1$. We observed each target at two different nod positions for a total integration time of 1--2 hours with each module. The data were obtained between August 2005 and June 2006.

We reduced the data using the {\it Spitzer} IRS S14 pipeline.\,\footnote{http://ssc.spitzer.caltech.edu/irs/dh/}  To ensure homogeneity throughout the sample, we reprocessed the raw data from our earlier results \citep{md07}, which had been processed by the S13 pipeline. We performed additional cleaning of the 2D spectra using the {\it SSC} utility {\sc irsclean}\,\footnote{http://ssc.spitzer.caltech.edu/archanaly/contributed/irsclean} to remove rogue pixels, by creating a mask for each Astronomical Observing Request (AOR). We use the rogue pixel mask associated with the observing cycle as a base and then identify additional deviant pixels in a {\it superdroop} file resulting from the averaging of all droop files within a single AOR.  We relied on differencing between the two nod positions to subtract the residual background. To do this, we median-combined all cycles for science images of a given SMG with the same module order (LL1 or LL2) and nod position. For each module order, we then subtracted the median of all science images at one nod position from that of the other nod position. 

\begin{figure*}
\centering
\includegraphics[scale=0.65, angle=0]{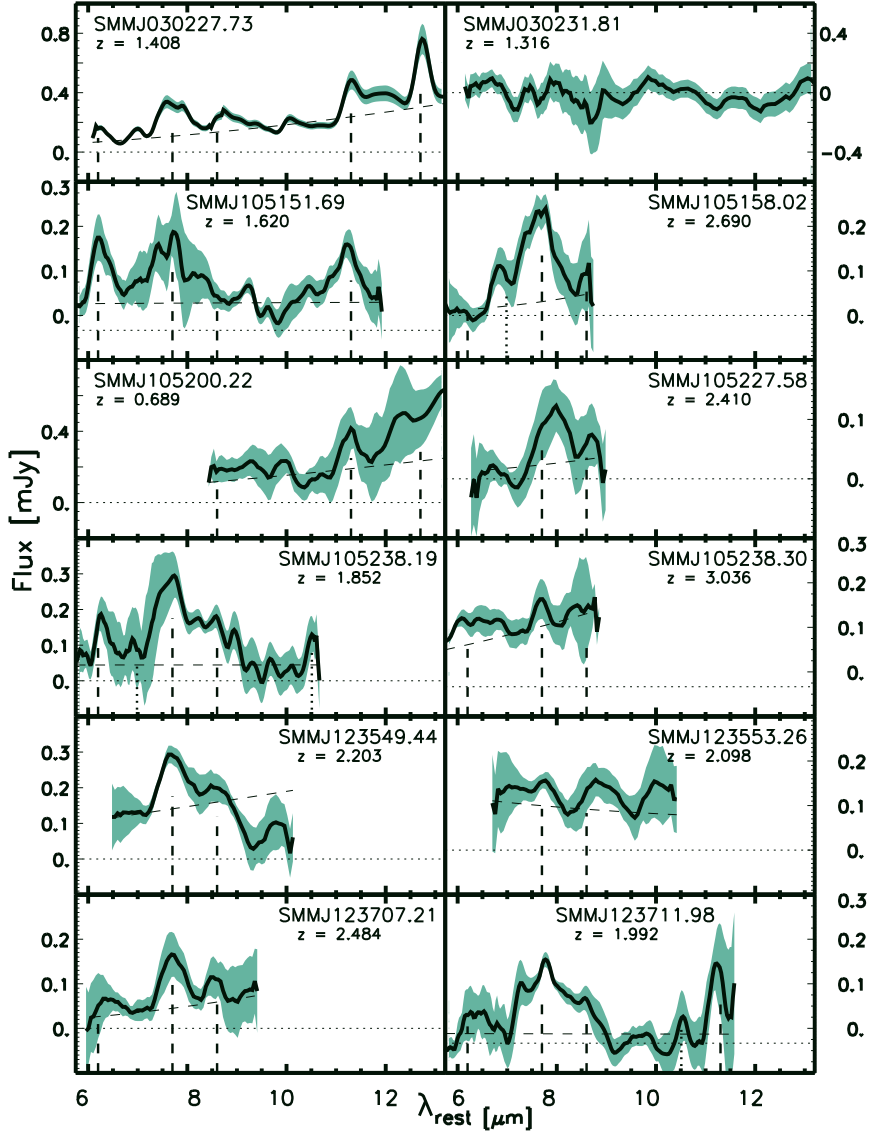}
\caption{Mid-IR spectra of all 24 SMGs in our sample redshifted to the rest-frame according to their rest-frame optical redshifts (C05) or their PAH-based redshift in the case of SMM\,J105151.69, SMM\,J105158.02, SMM\,J105227.58, SMM\,J221733.91 and SMM\,J221806.77 (see \S\ref{redshifts} for details). The spectra are smoothed to the instrument resolution ($\Delta\lambda \sim 0.3 \mu$m at $\lambda \sim 24 \mu$m). The continuum fit for each spectrum is denoted by a short-dashed line and the shading represents the flux uncertainty, as given by {\sc spice}. We do  not detect SMM\,J030231.81 at $z=1.316$. The locations of PAH features at 6.2, 7.7, 8.6, 11.3 and $12.7\,\mu$m are indicated by the dashed lines and these features are visible in a large fraction of our sample.  The position of tentative narrow line emission from [Ar{\sc ii}] ($6.99 \mu$m) and/or [S{\sc iv}] ($10.51 \mu$m) are shown by dotted lines in SMM\,J105158.02, SMM\,J105238.19, SMM\,J123711.98, SMM\,J123721.87, SMM\,J163639.01, SMM\,J221733.91 and SMM\,J221735.15.\label{spectra}
}
\end{figure*}

For each LL order, we used the SPitzer IRS Custom Extraction\,\footnote{http://ssc.spitzer.caltech.edu/postbcd/spice.html} ({\sc spice}) software to extract flux-calibrated 1D spectra from each 2D sky-subtracted median-combined nod. We used the {\it optimal extraction} mode included in {\sc spice}. Optimal extraction entails taking a weighted average of profile-normalized flux at each wavelength to increase the {\it S/N} in IRS observations of faint sources \citep{horne86, narron06}. Flux calibration is based on observations of a standard star during the same observing cycle as the science observations and has a 10\% uncertainty (see IRS Data Handbook\,\footnote{http://ssc.spitzer.caltech.edu/irs/dh/} for details). The {\sc spice}-extracted 1D spectra for the different nods within one LL order were then averaged together. We combined the orders using {\sc iraf} to produce a final spectrum for each target, excluding the noisy edges at $\lambda_{obs} \lesssim 14.2 \mu$m and $\lambda_{obs} \gtrsim 35 \mu$m. The resulting spectra are shown in Fig.~\ref{spectra}. 

%
%

%
%
\begin{figure*}
\centering
\includegraphics[scale=0.65, angle=0]{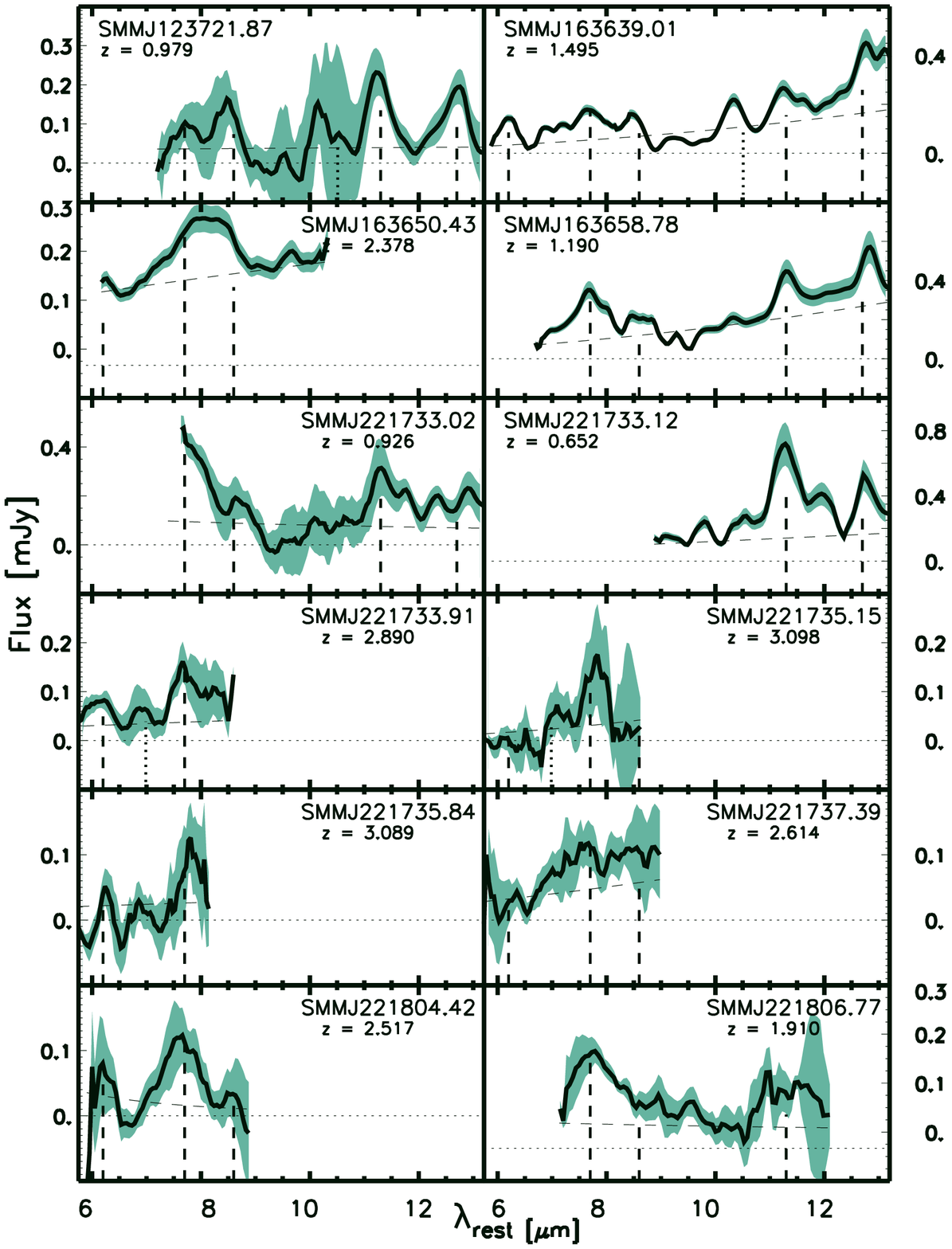}
\centerline{Fig.~2.\ --- Continued.}
\end{figure*}

%
%

\section{Analysis}\label{analysis}

The spectra in Fig.~\ref{spectra} show a range of emission and absorption features.  One of the key
advantages of our survey over previous studies is the availability of spectroscopic redshifts for the entire sample from C05.  
With spectroscopic redshifts in-hand, we have a-priori knowledge of the observed wavelengths where emission and absorption features are expected.
We therefore identified PAH features, silicate absorption centered at $\sim 9.7\mu$m and a number of possible forbidden narrow emission lines, such as [Ar{\sc ii}] ($6.99 \mu$m) and [S{\sc iv}] ($10.51 \mu$m). We see that a large fraction of our sample shows PAH emission features and we mark
these and the other spectral features on  Fig.~\ref{spectra}.

We fit our spectra to derive quantitative estimates of the strength of the emission and absorption features and determine their precise wavelengths. However, given our modest integration times (Table~\ref{Obstab}), we reach a signal to noise ({\it S/N}) of only 1--5 in the continuum emission at $24 \mu$m\footnote{{\it S/N} was determined within a $\lambda$-window of $\Delta\lambda \sim 1.4 \mu$m on the individual nod-subtracted 2D spectra}. At this {\it S/N} we found that intricate spectral fitting programs (e.g.\ {\sc pahfit}; \citealt{smith07}) are not appropriate.  We thus adopted a simple and straight-forward fitting approach which we
describe below.

Measurements of fluxes and equivalent widths (EW) of features in the spectra require that we first determine the continuum level for each spectrum before measuring the strength of each individual emission and absorption feature.
The resulting EW and flux measurements are extremely sensitive to the fitting method selected, particularly in the determination of the continuum level.  A number of alternative fitting methods have been used by different authors, making a fair comparison of PAH strengths in different samples a rather subjective and delicate issue \citep{sajina07}. In particular, the prevalence of interactive fitting methods has further complicated the reproducibility of mid-IR properties between samples. Hence in this work we have adopted a method of continuum and PAH fitting that ensures complete reproducibility. Moreover, in order to make fair comparisons between the SMGs in our sample and other low- and high-redshift sources, we have re-measured all the mid-IR features in all of these samples using our method. 
We have applied our method to local ULIRGs from the Bright {\it IRAS} sample studied by \citet{armus06, armus07} and \citet{desai07}, the low-redshift nuclear starburst-dominated galaxies presented by \citet{brandl06} and the mid-IR selected high-redshift sources of \citet{sajina07} and \citet{farrah08}. We note that within the context of PAH luminosities and EWs, we only include into our discussion those objects for which we could distinguish the PAH features. As we mentioned earlier, we have also included in our comparisons the 13 SMGs in the GOODS-North field studied by \citet{pope08}, nine of which have optical redshifts measured by C05 and thus naturally complement our sample.

%
%

\begin{figure*}
\centering
\includegraphics[scale=0.6, angle=270]{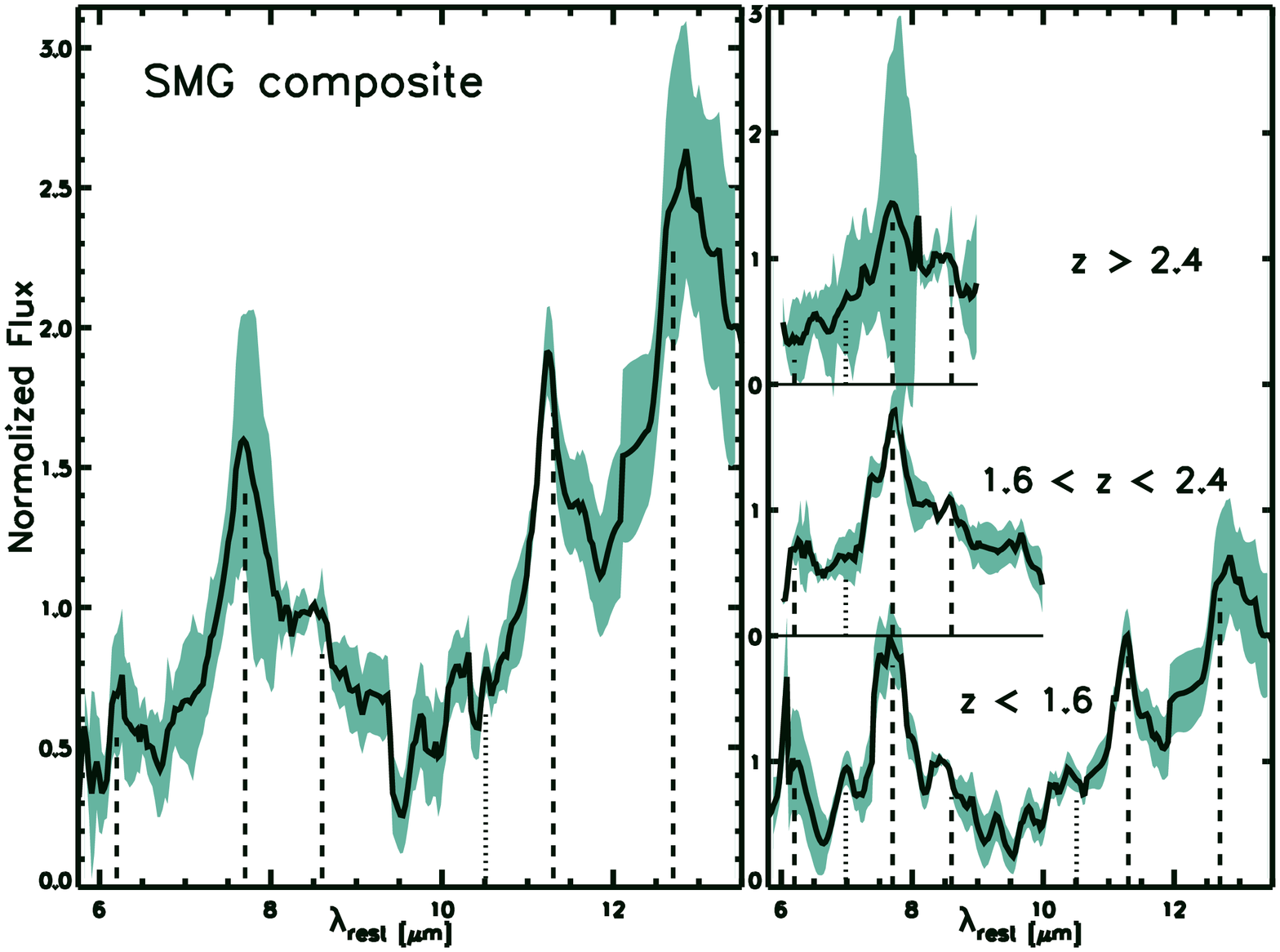}
\caption{({Left:}) The median-combined rest-frame composite spectrum of 22 SMGs detected in our sample, with the shaded area representing the 1-$\sigma$ sample standard deviation. We assume optical redshifts (C05), except for the five SMGs in which PAH features suggest an alternate redshift (see \S\ref{redshifts}).  The composite spectrum is dominated by strong PAH emission at  6.2, 7.7, 8.6, 11.3 and 12.7\,$\mu$m, with an underlying red continuum. All the spectra are  normalized by their flux at $\lambda = 8.5\,\mu$m in the rest-frame (see \S\ref{comp_construction} for details). ({Right:}) Composite spectra for subsets of SMGs in three separate redshift bins.  These demonstrate that at $\lambda > 10\mu$m our composite spectrum is dominated by sources at $z<1.6$. See \S\ref{SMGcomp} for discussion.
\label{SMGcomposite}}
\end{figure*}

The thermal mid-IR continuum, arising either from starburst or AGN activity can be described by a power-law, $S_{\nu} \sim \nu ^{-\alpha_{\rm MIR}}$, where the power-law index $\alpha_{\rm MIR}$ indicates the steepness of the spectrum. Flatter mid-IR continua with low power-law indices ($\alpha_{\rm MIR} \lesssim 0.5$) are usually associated with unobscured AGNs, where emission from the hot dust close to the AGN is readily observed. Steeper continua ($\alpha_{\rm MIR} \gtrsim 0.5$) are on the other hand associated with starburst activity, but can also be found in dust-obscured AGNs.  Hence we determined the power-law continuum index $\alpha_{\rm MIR}$ for the spectra in our sample to: (1) characterize the continuum slope and investigate the underlying energetics, and (2) to define the continuum level on which PAH emission features are superimposed. 

To determine $\alpha_{\rm MIR}$ we use the $\chi^2$ minimization {\sc idl} routine {\sc linfit} to obtain the best linear fit to each SMG spectrum in $\log(S_{\nu})$--$\log(\nu)$ space. We first obtain the best linear fit to each full spectrum and calculate the rms in the (data $-$ fit) residual spectrum. Then, we identify those wavelength regions in the original spectrum where the data points lie more than $1\times$ the rms away from this first linear fit. In this way, we automatically identify the wavelength regions where emission and absorption features are present. We then perform a second round of linear fitting, where we exclude these wavelength regions from the fit. We adopt the results of this second fit as our final continuum, with the influence of emission and absorption features minimized. The resulting fits to the continuum for each spectrum are shown in Fig.\,\ref{spectra}. 

Having determined the continuum, we search for the presence of PAH emission features at 6.2, 7.7, 8.6, 11.3 and 12.7$\mu$m.  Since our objects cover a range in redshifts of $z\sim0.65$--3.2, the wavelength coverage varies from $\sim9$--$22\mu$m for the lowest redshift SMG to $\sim$5--$9\mu$m for that with the highest redshift. Out of all detected SMGs in our sample, $\gtrsim$90\% cover of the $7.7\mu$m region of the spectrum, thus providing a large sample of 21 SMGs to investigate the spectral features within this important wavelength region. We define a wavelength window around each spectral feature and use the {\sc iraf} task {\sc splot} to fit the PAH features with individual Gaussians, where the center and FWHM are left as free parameters within the fitting window. We create a cursor file specifying the fitting windows and continuum level for each spectrum to automate the fitting process and avoid the lack of reproducibleness inherent to other interactive fitting tasks. The wavelength regions used to fit the 6.2, 7.7, 8.6, 11.3 and $12.7\mu$m PAH features were approximately 5.90--6.60, 7.10--8.3, 8.30--8.90, 11.00--11.60, 12.40--13.00\,$\mu$m, respectively. We adopt the {\sc splot} output uncertainties in the resulting PAH EWs and integrated fluxes, which account for the noise level present in the spectra. These results are presented in Table \ref{resultstab}.  

We take advantage of the PAH features in our spectra to determine mid-IR redshifts for the SMGs in our sample. For this we derive a redshift associated to each PAH feature from its observed central wavelength. The final PAH-based redshift for spectra displaying multiple PAH features corresponds to the weighted mean of the redshifts derived from the individual PAH features, where the weights are provided by the PAH fitting uncertainties. We present these results in \S\ref{redshifts}.

Finally, we quantify the strength of the silicate absorption. We define the optical depth as $\tau_{9.7 \mu m} = \log [S_{cont}/S_{obs}]$, where $S_{obs}$ is the flux level observed at $9.7 \mu$m and $S_{cont}$, the interpolated flux of continuum in the absence of any absorption.  We measure the latter from our power law continuum fit.  In \S\ref{tau} we present the resulting values for the SMGs in our sample that provide sufficient wavelength coverage for a reliable determination of the continuum.

%
%

\subsection{Composite Spectra}\label{comp_construction}

In the interest of obtaining a representative spectrum for the sample as a whole, with a higher {\it S/N} to enable identification of fainter underlying features, we also median-combined the individual rest-frame spectra of the SMGs in Fig.\,\ref{spectra} into a composite spectrum.  The full collection of individual spectra in our sample cover the rest-frame wavelength range $\lambda \sim 5$--22\,$\mu$m. To avoid being overly biased towards the mid-IR properties of SMGs within a narrower redshift range, we only include in the composite spectrum wavelengths covered in {\it at least} five SMGs. The individual spectra in Fig.~\ref{spectra} also display a significant spread in rest-frame continuum level with $S_{9\mu m} \sim 0.03$--0.2\,mJy and so we normalize them by their flux at a rest-frame wavelength of 8.5$\mu$m. Only the spectrum of SMM\,J221733.12 does not cover this wavelength and is thus excluded from the composite. The final composite spectrum thus includes 22 detected SMGs and is shown in Fig.~\ref{SMGcomposite}. The sample standard deviation of the composite spectrum increases towards longer wavelengths, in part reflecting the fact that fewer individual spectra contribute at $\lambda_{rest} \gtrsim 10 \mu$m.  To better demonstrate the redshift ranges which contribute to the different wavelength regions in the composite (and their variations) we divide our sample into three different redshift bins and construct composite spectra of low-, intermediate- and high-redshift SMGs (see Fig.~\ref{SMGcomposite}) and discuss these in \S\ref{SMGcomp}. 

We also take advantage of the range in properties covered by our sample (see Fig.~\ref{C05}) to investigate mid-IR spectral properties of different SMG sub-populations. We follow the method described above to construct composite spectra of subsamples of SMGs based on $24 \mu$m- and radio brightness to search for trends within our sample. However, composite spectra based on apparent brightness  mix sources with different intrinsic luminosities at different redshifts. We attempt to minimize this bias by first defining luminosity-complete subsamples of SMGs, then restrict the SMGs included in each composite to those that fall within a redshift range where we are close to complete in terms of mid-IR and radio luminosity. At the median redshift of our sample, $<\! z\! > \sim 2$, the observed $24 \mu$m flux corresponds to redshifted rest-frame $8 \mu$m emission. Therefore, we define  luminosity-complete subsamples of SMGs by using the rest-frame luminosities at $8 \mu$m and at 1.4\,GHz.\footnote{L$_{1.4 GHz} = 4\pi D_L^2 S_{1.4 GHz} (1+z)^{-(\alpha + 1)}$\,W\,Hz$^{-1}$, assuming $\alpha = -0.8$, the average spectral index for star-forming galaxies (e.g., \citealt{yun01})} . To build the $24 \mu$m-composites we restrict the parent sample to SMGs with L$_{8 \mu m} \gtrsim 10^{32}$ erg s$^{-1}$. For the radio composite, we include SMGs with L$_{1.4 GHz} \gtrsim 2 \times 10^{24}$ erg\,s$^{-1}$\,Hz$^{-1}$. 
As a result, low-redshift SMGs with the lowest mid-IR and radio luminosities in the full sample, even though they have the highest apparent fluxes, are excluded from the $24 \mu$m- and radio-brightness composites.  Results for the various composite spectra from \S3.1 are presented in Table~\ref{compositestab}.  We discuss these composites in \S\ref{RadioAND24umcomp}.

%
%

\section{Results}\label{results}

Out of the 24 SMGs in our sample shown in Fig.~\ref{spectra}, we detect PAH emission and continuum in 15 SMGs, PAH emission and  no continuum in four, continuum but no PAH emission in another four and neither continuum nor PAH emission in only a single SMG (SMM\,J030231.81 at $z=1.316$).  Hence, our sample contains 19 SMGs with PAH emission and 19 with detectable continuum emission.
In addition, marginal detections of [Ar{\sc ii}] ($6.99 \mu$m) and [S{\sc iv}] ($10.51 \mu$m) narrow-line emission were found in several individual SMGs (see Fig.\,\ref{spectra}). 
 
%
%
\begin{figure}
\plotone{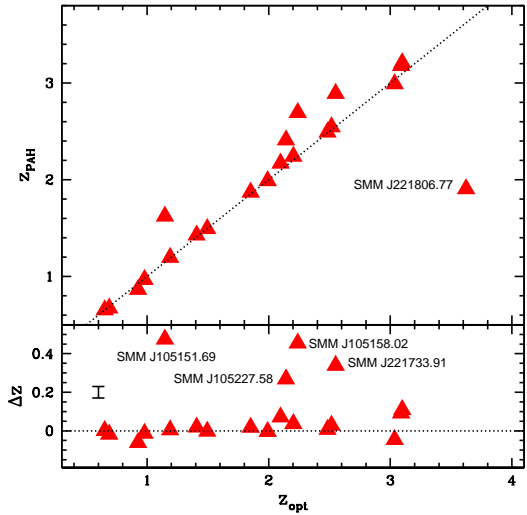}
\caption{A comparison of the  PAH-derived redshifts with the rest-frame UV redshifts for the SMGs in our sample (C05).  PAH-based redshifts, with a typical uncertainty $\Delta z = 0.03$ agree well with the optical redshifts determined by C05, except for five out of the 23 detected SMGs, namely SMM\,J105151.69, SMM\,J105227.58, SMM\,J105158.02, SMM\,J221733.91 and SMM\,J221806.77. These SMGs have multiple PAH features that suggest a redshift significantly different from the optical one. Four out of these five SMGs have mid-IR spectra displaying one particularly prominent feature that we associate with the $7.7 \mu$m PAH line. The presence of additional fainter features confirm the validity of this assumption.  \label{zfig}}
\end{figure}

%
%

\subsection{PAH-based Redshifts}\label{redshifts}

We have rest-frame UV redshifts for all the SMGs in our sample from optical spectroscopy using the Low-Resolution Imaging Spectrograph on Keck\,II (C05). These spectroscopic redshifts allowed us to identify confidently faint emission and absorption features in the mid-IR spectra of individual SMGs. Nevertheless, with 19 SMGs displaying prominent PAH emission, we can determine a mid-IR redshift for these SMGs based on the location of the redshifted PAH features and compare these to the rest-frame UV redshifts. PAH-based redshifts ($z_{PAH}$) for our sample are shown in Table~\ref{resultstab} and are compared to the redshifts from C05 in Fig.~\ref{zfig}.  We note that PAH-derived redshifts may have large uncertainties ($\Delta z \sim 0.01-0.05$) arising from the intrinsically large width of PAH features (FWHM$\gtrsim 10^3$\,km\,s$^{-1}$). Redshifts based on rest-frame UV features are thus more precisely determined. However, we note that because the optical emission in these systems does not trace the bolometric emission well, there are opportunities for these rest-frame UV redshifts to be erroneous if the optical spectroscopic IDs are spatially offset from the radio counterpart.  Equally, the large IRS aperture may result in mid-IR bright sources unrelated to the SMG to dominate the spectrum. 

The comparison in Fig.~\ref{zfig} demonstrates that the IRS spectra of the majority of the SMGs have features consistent with the rest-frame UV redshifts.  However, there are five SMGs  where the features suggest an alternate redshift (see Table\,\ref{resultstab}) that is significantly different from the rest-frame UV redshift. For the five SMGs with discrepant redshifts, the reasons for the disagreement are in all cases due
to ambiguities and line misidentifications resulting from low
{\it S/N} features in the UV spectra, and so we discuss these individually.
We note that these corrections to the redshifts have not
significantly changed the interpretation of these sources.\footnote{Changes in redshift at the 10--20\%
level at $z\sim 2$ do not result in significant changes to the luminosities or
temperatures beyond the associated calibration errors.}
We also note that one of the new PAH redshifts, SMM\,J105158.02,
strengthens the $z\sim2.7$ SMG association discussed in \citet{blain04}.

In the case of SMM\,J221733.91, an initial indication of a redshift around
$z\sim2.5$ from absorption features in a low {\it S/N} spectrum was apparently verified
by a weak H$\alpha$ line at $z=2.555$ (\citealt{swinbank04}). However, a deeper UV-spectrum, doubling
the {\it S/N}, revealed an unambiguous $z=2.865$ from multiple absorption
features, consistent with the $z=2.89$ found from the IRS spectrum.

Similarly with SMM\,J105227.58, the original noisy UV absorption
line spectrum was somewhat ambiguous but suggested $z=2.142$. Subsequent
deeper spectroscopy has revealed this source to be $z=2.470$ from weak
Ly$\alpha$ in emission as well as C{\sc iv}{1549} absorption and emission, consistent with the PAH redshift within fitting uncertainties.

SMM\,J221806.77 was identified in C05 at $z=3.623$ from a single weak emission
line detection in a relatively sky-free spectral region and corroborated by a
red continuum dropping out in the $B$-band.
The PAH-based redshift $z=1.91$ suggests that the continuum of this source is heavily
reddened, and that the rest-frame UV emission feature is either noise or a superposed
line detection from a different galaxy.

SMM\,J105158.02 was identified at $z\sim2.2$ first in the UV through
various low {\it S/N} absorption features. An apparent H$\alpha$
detection consistent with $z=2.239$ was proposed as the true redshift
(\citealt{swinbank04}). Re-analysis of more recent, deeper UV-spectra does reveal a suite of absorption
features consistent with $z=2.694$ (Si{\sc ii}\,{1260}, O{\sc i}/Si{\sc ii}\,{1303},
C{\sc ii}\,{1335}).

Finally, for SMM\,J105151.69, an apparent emission line proposed as
[O{\sc ii}]\,{3727} is not reproduced in later Keck spectra taken on this galaxy
at a different position angle, and it has now been identified as an
emission line galaxy slightly offset from the radio position.
The PAH-based redshift of $z=1.62$ is consistent with UV-spectra
[C{\sc iii}]\,{1909} absorption features and Mg{\sc ii}\,{2800} emission features.

%
%
\begin{figure}
\plotone{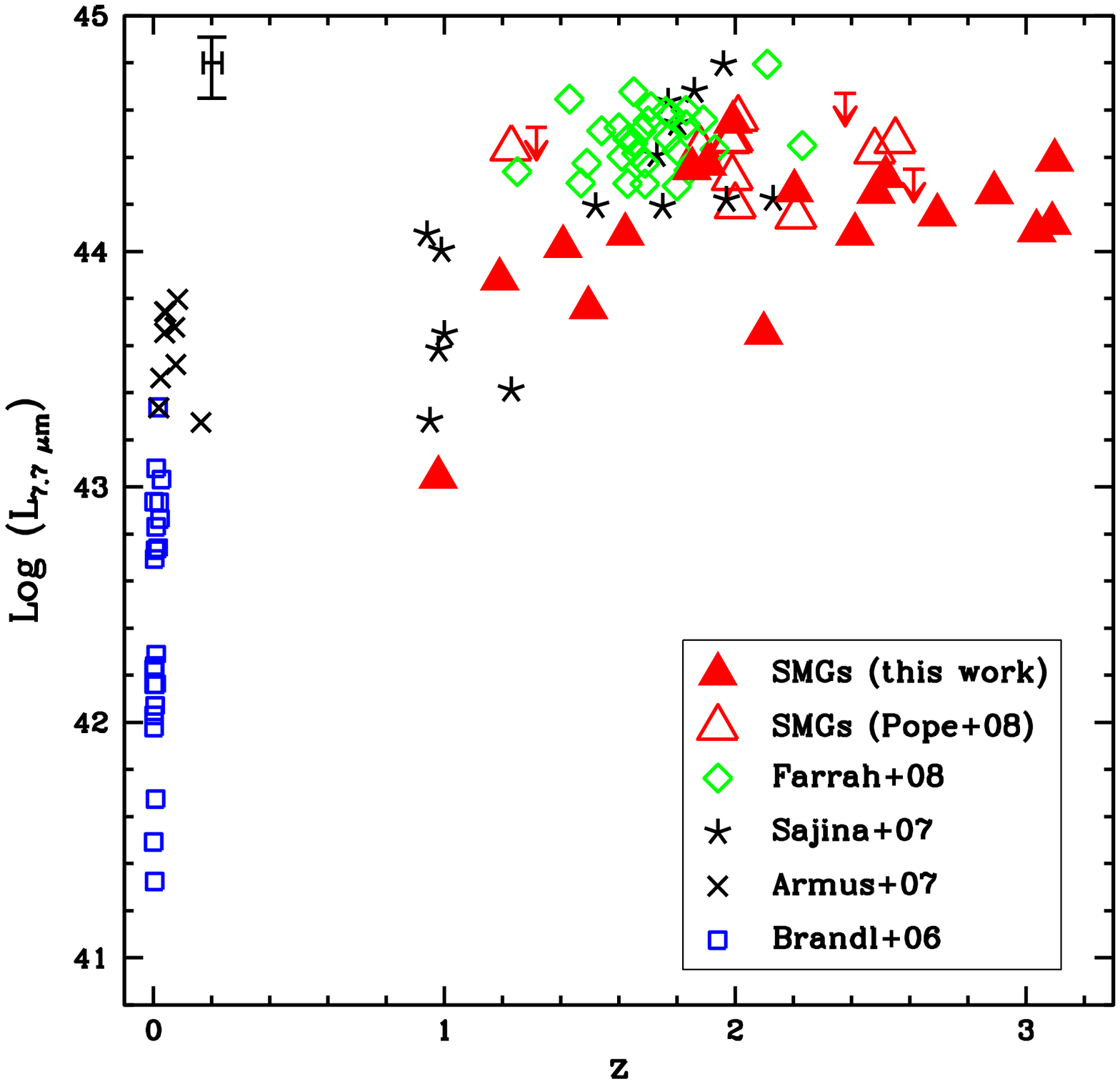}
\caption{The $7.7\mu m$-PAH luminosities as a function of redshift for the SMGs in our sample.  In addition we plot the SMGs from \citet{pope08}. For comparison we also plot aperture-corrected low-redshift nuclear starburst-dominated galaxies \citep{brandl06}, local ULIRGs from the bright {\it IRAS} sample (\citealt{armus06, armus07}) and high-redshift mid-IR-selected sources \citep{sajina07, farrah08}. We see that SMGs are among the most PAH-luminous objects in the universe, with PAH features up to $1000\times$ and $10\times$ more luminous than local starbursts and ULIRGs, respectively. Considering that PAH emission is associated to strong star-forming activity, this suggests that the SMG population comprises some of the most vigorous star formation events ever observed.
See \S\ref{pahs} for discussion. \label{pahVSz}}
\end{figure}

%
%

\subsection{PAH Luminosities and PAH ratios}\label{pahs}

The $7.7 \mu$m PAH feature is the most prominent in the mid-IR spectra of star-forming galaxies and is thus particularly useful in studying star formation in faint sources at high redshift. We derive PAH luminosities from the integrated line fluxes presented in Table~\ref{resultstab} and plot these in Fig.~\ref{pahVSz}. We find the bulk of $7.7 \mu$m PAH luminosities for SMGs in the range $\sim 10^{44}$--10$^{45}$\,erg\,s$^{-1}$ and hence that, as expected from their far-IR luminosities, SMGs have PAH luminosities that exceed those of  low-redshift ULIRGs \citep{armus07}\footnote{We include NGC\,6240, Mrk\,1014, UGC\,5101, Arp\,220, FSC\,05189$-$2524, FSC\,12112$-$0305, Mrk\,273, FSC\,14348$-$1447, FSC\,22491$-$1808 from \citet{armus07}.} and local nuclear starbursts \citep{brandl06} by factors of up to 10 and 1000, respectively.  These PAH luminosities indicate that $\sim 0.5$--5\% of the bolometric luminosity in SMGs escapes in the form of PAH emission. 

From Fig.~\ref{pahVSz} we can see that the range in PAH luminosities for SMGs covers that occupied by the mid-IR selected sources in \citet{sajina07} and \citet{farrah08}. The sample from \citet{sajina07}, based on mid-IR colors and bright  $24\mu m$ fluxes ($S_{24 \mu m} \gtrsim 0.9$ mJy), has been shown to comprise mostly composite objects, dominated by AGN activity ($\sim 75\%$; \citealt{sajina07}). Roughly $25\%$ of their sample display strong PAH features, which explains the large PAH luminosities displayed in Fig.~\ref{pahVSz}, ranging from values comparable to low-redshift ULIRGs to those shared with SMGs.  Even though these objects have similar PAH luminosities to that of SMGs, their PAH EWs are on average lower, with SMGs displaying a  median $6.2\mu$m PAH EW three times larger and a median $7.7\mu$m PAH EW larger by a factor of $\sim1.5$. Probing slightly lower $24\mu$m fluxes ($S_{24 \mu m} \gtrsim 0.5$ mJy), \citet{farrah08} also select bright mid-IR sources at high redshift, but with an additional criterium of stellar-dominated near-IR SEDs. These objects concentrate around $z\sim1.7$ and Fig.~\ref{pahVSz} shows that they display similar PAH luminosities to SMGs and to the most PAH-luminous objects in the sample from \citet{sajina07}. The mid-IR selected objects in \citet{farrah08}, in contrast to those in \citet{sajina07}, have PAH EWs similar to SMGs and more than a factor of two larger than those of the mid-IR bright objects in \citet{sajina07}, indicating a reduced AGN contribution to their  mid-IR emission compared to the latter sample (see \citealt{farrah08}). We discuss further the comparison of SMGs and other galaxy populations in \S\ref{discuss}.

Different PAH emission features arise from distinct bending/vibrational modes of the PAH molecules \citep{draine07}. The different modes can be enhanced relative to each other as a result of varying PAH ionization state and PAH size distribution. Exposure to a more energetic radiation field may strip PAH molecules of peripheral H-atoms. A decrease in the fraction of neutral PAHs would diminish the strength of the 8.6, 11.3 and $12.7 \mu$m PAH features, which are produced by the in-plane ($8.6\mu$m) and out-of-plane (11.3, $12.7 \mu$m) C-H bending mode. As a result, the 6.2 and $7.7 \mu$m PAH features, which are thought to be produced by C-C skeleton vibration \citep{draine07}, would appear to dominate. Within this context, the relative strength of PAH features can be used to probe the energetics of the underlying radiation field (e.g., \citealt{galliano08}). 
The size distribution of PAH molecules may also affect the observed relative strength of  features:  smaller PAHs (N$_{C} \lesssim 10^3$) tend to emit more strongly at 6.2 and $7.7 \mu$m \citep{allamandola89}. In the hottest regions close to an AGN, sublimation of these smaller PAH molecules may result in a supression of short-wavelength PAH emission. Dust extinction has also been shown to play an important role in explaining the variation in the relative strength of PAH features. In particular, strong extinction can increase the 7.7/6.2 and 7.7/8.6 PAH ratios \citep{rigopoulou99} due to an ice absorption feature at $\sim 6\mu$m (which reduces the 6.2$\mu$m PAH emission, \citealt{spoon02}) and broad silicate absorption at $\sim 9.7\mu$m, where the former reduces the strength of 6.2$\mu$m PAH feature and the latter that at 8.6$\mu$m.

\begin{figure}
\plotone{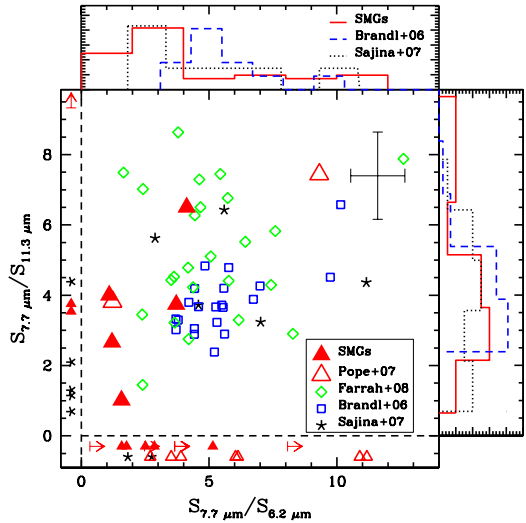}
\caption{Distribution of $7.7/6.2$ and $7.7/11.3$ PAH ratios within the SMG population. Objects for which we have only measurable  7.7/6.2 PAH ratios are plotted at $S_{7.7}/S_{11.3} <0$ and those for which we only have 7.7/11.3 measurements are plotted at $S_{7.7}/S_{6.2} <0$.  We also include PAH ratios measured for a sample of low-redshift nuclear starbursts \citep{brandl06}, and high-redshift mid-IR selected sources (\citealt{sajina07, farrah08}). SMGs share similar $7.7/11.3$ PAH ratios with these populations, but their distribution in $7.7/6.2$ PAH ratios is skewed towards values that are in general lower than that of low-redshift starbursts. We suggest that this difference in $7.7/6.2$ PAH ratio may arise due to a difference in extinction (see \S\ref{pahs} and \S\ref{dust} for discussion).
\label{PAHratios_dist}}
\end{figure}

We consider the flux ratios in the 6.2, 7.7 and $11.3 \mu$m PAH features in our sample to investigate the energetics and extinction in the star-forming environments within SMGs. We plot these in Fig.\,\ref{PAHratios_dist} and find that SMGs have median 7.7/11.3 and 7.7/6.2 PAH ratios of $3.8\pm 1.6$ and $2.5 \pm 1.4$, respectively. The median 7.7/11.3 PAH ratio is similar to that of low-redshift nuclear starbursts, $S_{7.7}/S_{11.3}=3.7$. However, the distribution in values for the  7.7/11.3 PAH ratio in  SMGs is very broad, extending over a similar range to that of the high-redshift $24\mu$m-selected sources studied by \citet{sajina07} and \citet{farrah08}. Such diversity in PAH ratios is also present in local ULIRGs (e.g., \citealt{rigopoulou99,peeters04}), indicating that diverse radiation fields are found amongst both low- and high-redshift IR-luminous sources. We note that the need for measurable $11.3\mu$m emission means that this comparison only includes SMGs at $z\lesssim1.6$ (Fig.~\ref{SMGcomposite}).

In Fig.\,\ref{PAHratios_dist} we illustrate that the bulk of SMGs have 7.7/6.2 PAH ratios lower than the median value found for local nuclear starbursts ($S_{7.7}/S_{6.2}=2.5$ for SMGs vs. $5.3$ for the starbursts).  This suggests that an intrinsic difference in either radiation field or extinction may exist between SMGs and low-redshift nuclear starbursts, which result in a suppressed $6.2 \mu$m PAH feature and a relatively prominent $7.7 \mu$m PAH feature \citep{rigopoulou99}.  As we discuss later, we believe that reduced extinction may be responsible for this behaviour in SMGs.

%
%

\subsection{Continuum Slopes}\label{alpha}

The continuum power-law index $\alpha_{\rm MIR}$ characterizes the steepness of the mid-IR continuum and requires adequate wavelength coverage to be measured reliably. In Table~\ref{signtab} we present the $\alpha_{\rm MIR}$ values for the SMGs whose spectra cover a wavelength range that extends longwards of the silicate absorption and that we thus consider reliable.  We find that the SMGs in our sample have a median mid-IR power-law index of $<\alpha_{\rm MIR}> = 1.05$, typical of the continuum found for star-forming regions.  However, the individual spectra span a range of power-law indices $\alpha_{\rm MIR} \sim -0.8$--2.5. Considering the uncertainties in the individual fits, with a median average of $<\Delta\alpha_{\rm MIR}>\sim0.3$, this range in power-law indices suggests a real diversity in the mid-IR continuum slopes within the sample. Power-law indices in $\sim 60$\% of the SMGs in our sample are consistent with those found in star forming regions  ($\alpha_{\rm MIR} \gtrsim 0.5$), while the remaining $\sim40$\% have  $\alpha_{\rm MIR} \lesssim 0.5$, more  characteristic of AGN-dominated regions (e.g., \citealt{deo07}).  

%
%

For a subset of our sample we can also  characterize the continuum shape from the $S_{6\mu m}/S_{12\mu m}$ color, which is measured at wavelengths devoid of substantial contamination from PAH emission. We indicate these results in Table~\ref{signtab}. At $6 \mu$m, the continuum traces thermal emission from hot (T$_{dust} \sim 500$\,K) dust, while at $12 \mu$m, the continuum is dominated by dust heated to lower temperatures (T$_{dust} \sim 250$\,K). We find that $\sim 60$\% of the SMGs for which we can determine mid-IR colors show red continua, with $S_{6\mu m}/S_{12\mu m} \lesssim 0.3$ and $\alpha_{\rm MIR} \gtrsim 0.5$. These resemble the colors of
low-redshift star-forming galaxies and 
the nuclear starburst galaxies presented by \citet{brandl06}, with $S_{6\mu m}/S_{12\mu m} \sim 0.18$--0.53. The remaining 40\% of this subsample have bluer continua ($S_{6\mu m}/S_{12\mu m}) \sim$ 0.8--1.1 and $\alpha_{\rm MIR} \lesssim 0.5$), more typical of the high-redshift mid-IR selected sources of \citet{sajina07} with $S_{6\mu m}/S_{12\mu m} \gtrsim 0.3$. 

Taken together these results  suggests that the mid-IR continuum emission in $\sim 60\%$ of the SMGs in our sample has the spectral characteristics associated with star forming systems locally.  The remaining third of the SMGs show more AGN-like mid-IR continua, although these frequently also display strong PAH emission.

%
%

\subsection{Obscuration as Measured by Silicate Absorption}\label{tau}

One of the clearest indicators of absorption in the mid-IR arises from intervening amorphous silicate dust grains.  Silicate absorption from the stretching of the Si-O bonds is centered around $9.7 \mu$m and is the main absorption feature within the wavelength range considered in this work. It is measurable in SMGs at $z\lesssim2.0$, corresponding to roughly half of our sample. 

We use the strength of the silicate-absorption feature as a measure of the obscuration along the line of sight to the warm continuum emission, $\tau_{9.7\mu m}$. We note that this definition of optical depth assumes that the obscuring material lies in a foreground screen, which is likely to be too naive a model for the complex mix of absorption and emission in SMGs.

We calculate $\tau_{9.7\mu m}$ for the individual SMGs, but stress that since we only detect faint continuum with {\it S/N}\,$ \lesssim 3$ for a number of SMGs in our sample, the values we derive for the strength of the silicate absorption are likely lower limits (see Table~\ref{resultstab}). We find a range of optical depths limits $\tau_{9.7 \mu m} \sim 0.05$--1.2, with a median value of $\tau_{9.7\mu m} =0.31$.\footnote{Including the SMGs with redshifts from C05 from \citet{pope08} extends this range to $\tau_{9.7 \mu m} \sim 1.4$.}  For the composite spectrum, which benefits from an improved {\it S/N}, we find an optical depth of $\tau_{9.7 \mu m} = 0.44 \pm 0.06$.

\begin{figure}
\centering
\includegraphics[scale=0.4, angle=0]{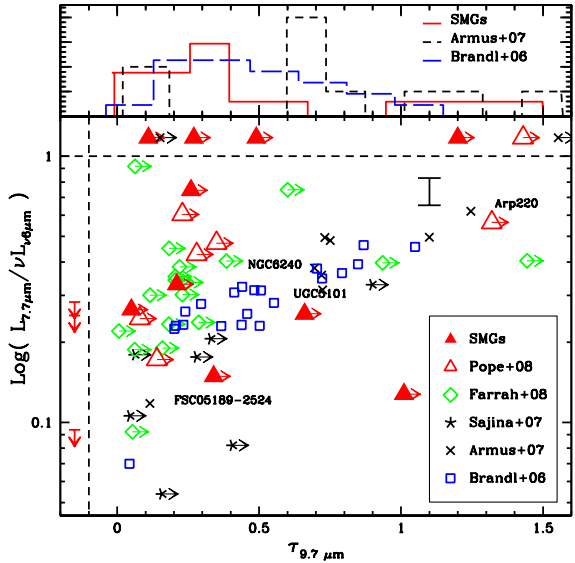}
\caption{A plot of the $7.7 \mu$m PAH luminosity relative to the 6-$\mu$m continuum luminosity as a function of silicate optical depth. 
We include optical depth measurements only for those SMGs that provide sufficient wavelength coverage for reasonable continuum estimates: 10 SMGs from our sample and seven SMGs from \citet{pope08}.  Optical depths for local nuclear starbursts \citep{brandl06}, local ULIRGs \citep{armus07} and high-redshift $24\mu$m-selected sources \citep{sajina07, farrah08} are also shown for comparison. Accurate measurements of optical depths require careful determination of the unobscured continuum. We thus quote the values for high-redshift sources displaying low S/N continuum detection at $\lambda=9.7\mu$m as lower limits. Optical depths measurements for objects for which we do not have continuum measurements at $S_{6\mu m}$ are plotted at $Log(L_{7.7}/L_{6}) >1$; upper limits for $Log(L_{7.7}/L_{6})$ in SMGs with no $\tau_{9.7\mu m}$ measurements are shown at $\tau_{9.7\mu m}<-0.1$. The error bar displays the median uncertainty in the SMG PAH-to-continuum ratios.   The histogram at the top shows the  distributions for SMGs and the two low-redshift samples. SMGs display $\tau_{9.7\mu m}$ limits lower than local ULIRGs and the most obscured starbursts, but have proportionally stronger 7.7$\mu$m PAH emission. 
\label{PAHcontTau}}
\end{figure}

In Fig.~\ref{PAHcontTau} we present optical depths ($\tau_{9.7 \mu m}$) as a function of the $7.7\mu$m PAH luminosity relative to the hot-dust continuum, $L_{7.7\mu m}/L_{6\mu m}$. We compare SMGs to low-redshift starbursts and ULIRGs, and to high-redshift mid-IR selected sources. We stress that the IRS observations of the local ULIRGs contain 80--90\% of the star-formation activity in these systems and hence the differences in apparent properties are not due to comparing galaxy-integrated measurements for the SMGs to measurements of just the nuclear component in local systems \citep{armus07}. 

The PAH-to-continuum luminosity ratio closely resembles a measure of EW as it probes the strength of the PAH feature with respect to the continuum at neighboring wavelengths.  The distribution in PAH-to-continuum values for SMGs is on average similar to that of mid-IR selected high-redshift sources of \citet{farrah08}, but higher than that of the mid-IR bright sources of \citet{sajina07}. This is in agreement with differences in their PAH EWs as discussed in \S\ref{pahs}, suggesting an increased AGN contribution to the mid-IR in the latter sample.

Fig.~\ref{PAHcontTau} also shows that at low redshifts, starburst galaxies with the highest PAH-to-continuum ratios display the highest optical depths, suggesting that the most active starbursts are also the most obscured, with optical depths close to that of local ULIRGs, $<\tau_{9.7\mu m}> = 0.7$ (see also \citealt{desai07}). SMGs lie off this trend, displaying a distribution in PAH-to-continuum values similar to low-redshift starbursts ($L_{7.7}/L_{6} \sim  0.1$--0.8), but maintaining on average $\tau_{9.7 \mu m}$ limits that are typically less than those seen in low-redshift ULIRGs and in the most obscured local nuclear starbursts. This difference in distribution suggests that SMGs have lower extinction along the line of sight to their warm continuum emission sources with respect to these low-redshift samples. Such a difference in distribution of silicate optical depths with respect to the local samples is also seen in the mid-IR selected objects of \citet{farrah08}.  

%
%

%
%
\begin{figure*}
\plottwo{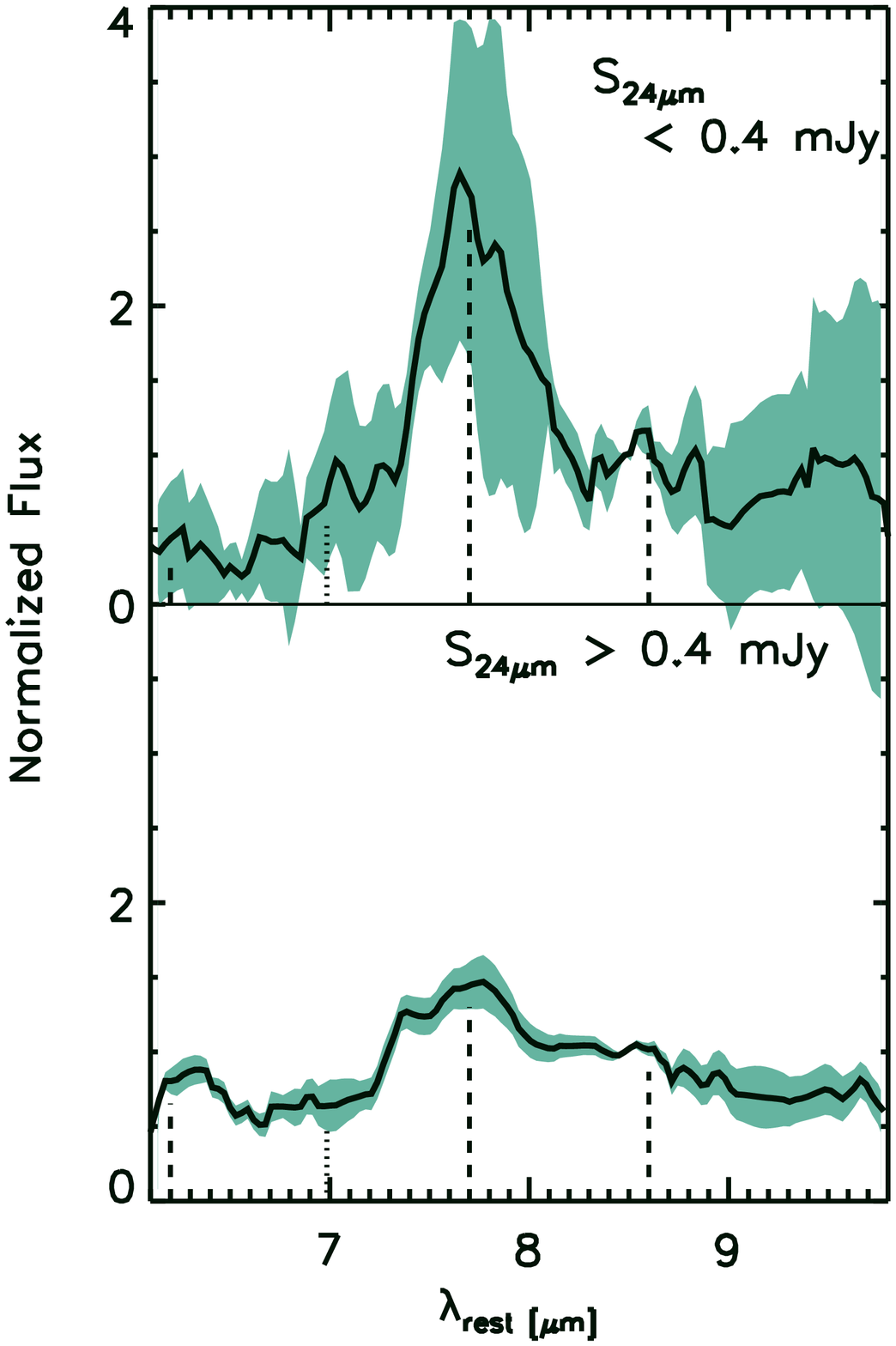}{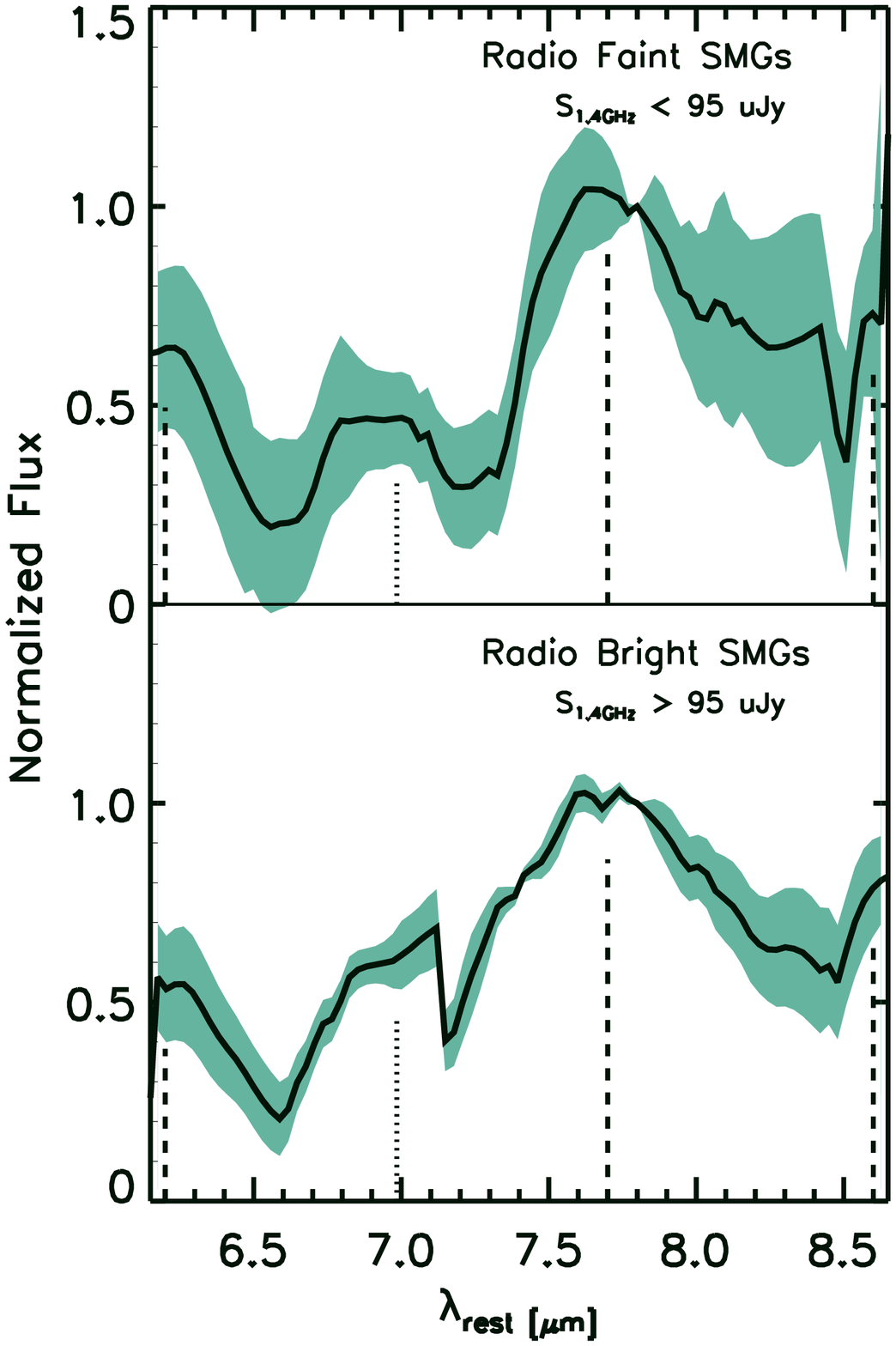}
\caption{({Left:})  The composite spectra of the seven $24 \mu$m-bright SMGs within the $L_{8\mu m}$-complete sample (bottom) and of the seven $24 \mu$m-faint SMGs (top).  ({Right:})  Composite spectra of the five radio-bright SMGs within the $L_{1.4 GHz}$-complete sample (bottom) and of the five radio-faint SMGs (top). We normalize the spectra of the individual SMGs that comprise the radio composites at $\lambda = 7.8 \mu$m. The $24\mu$m-faint composite is significantly more PAH-dominated than the $24\mu$m-bright composite, while the latter displays a stronger continuum. In the case of the radio-composites, the radio-bright sample appears to display a broader $7.7\mu$m PAH feature over a slightly more prominent continuum.  However, characterization of the radio-composite spectra does not reveal any significant difference within our measurement uncertainties (see Table\,\ref{compositestab}).
\label{Radio_24composites}}
\end{figure*}

%
%

\subsection{Composite Spectra}\label{composites}

%
%

\subsubsection{The SMG Spectrum}\label{SMGcomp}

With clear PAH features in $\gtrsim 80\%$ of the SMGs in our sample, the composite spectrum in Fig.\,\ref{SMGcomposite} displays prominent PAH emission. Taking advantage of the wide redshift range covered by our sample, we explore the composite spectra of SMGs in three different redshift bins: a low-redshift bin ($z \leq 1.6$), an intermediate-redshift bin ($z=1.6$--2.4) and a high-redshift bin ($z \geq 2.4$); see Fig.\,\ref{SMGcomposite}. The SMGs in the low-redshift bin dominate the composite spectrum at $\lambda \gtrsim 10 \mu$m. Therefore, we can track the evolution of the mid-IR spectral properties of SMGs at $\lambda \lesssim 10 \mu$m using our sample. 

All three redshift composites in Fig.\,\ref{SMGcomposite} display strong $7.7 \mu$m PAH emission and the $8.6 \mu$m feature also remains prominent.  With increasing redshift we observe a broadening of the $7.7 \mu$m PAH feature and a weakening of the $6.2 \mu$m PAH feature. The increase in scatter around the median composite spectrum indicates either that a more diverse range in spectral properties is present at the higher redshifts covered by our sample or poorer {\it S/N}. However, we note that the three composites display an increase in continuum level with redshift. Within the context discussed in \S\ref{pahs}, the weakening of the observed $6.2\mu $m-PAH emission feature may be an indication that SMGs at higher redshifts suffer from increased extinction, hampering detection of PAH emission features at the shorter wavelengths, or from a hardening of the radiation field the PAH molecules encounter (see also \citealt{watabe09}).

%
%

\subsubsection{Investigating Potential Sample Selection Biases}\label{RadioAND24umcomp}

To investigate how representative our conclusions are for the whole SMG population we need to investigate the variation of mid-IR properties of SMGs with both apparent 24$\mu$m and radio flux. Considering that one of our selection criteria for building the SMG sample relied on choosing objects with higher (estimated) $24\mu$m fluxes, we probe for differences between 24$\mu$m-bright and 24$\mu$m-faint SMGs (in the observed frame) to understand how representative our sample is of the SMG parent population with the faintest 24-$\mu$m counterparts. Using the SMGs within the $L_{8 \mu m}$-complete sample described in \S\ref{comp_construction}, we construct mid-IR composite spectra based on $24 \mu$m-brightness: SMGs with $S_{24 \mu m} \gtrsim 0.4$\,mJy comprise the $24\mu$m--bright composite, while those with $S_{24 \mu m} \lesssim 0.4$\,mJy make up the $24\mu$m--faint composite. We plot these composites in Fig.\,\ref{Radio_24composites}.  

We see strong $7.7 \mu$m PAH emission in the $24\mu$m--faint composite with EW$_{7.7\mu m}=2.48\pm0.35$, while a greater contribution from hot dust continuum in the $24\mu$m--bright composite results in a significantly lower $7.7\mu$m PAH EW, EW$_{7.7\mu m}=0.99\pm0.14$. Thus it appears that the strong PAH emission we see in our SMG sample at $z\sim 2$ is not likely to be a result of our $24 \mu$m-criterium in defining our initial target sample: the composite spectrum of SMGs with faint $24\mu$m-fluxes also displays strong PAH features, even stronger than those in the $24\mu$m-bright sample. The latter point even suggests that the prevalence of sources with strong continuum emission (and hence the contribution from AGN or highly obscured starbursts) is likely to be higher in our sample than in the general SMG population.

We use the range in radio fluxes covered by our sample (see Fig.~\ref{C05}) to investigate possible intrinsic trends in mid-IR properties associated with radio-brightness. We seek to address potential differences between the C05 sample of radio-identified SMGs and the $\lesssim 30$\% of $S_{850\mu \rm m}\gtrsim 5$\,mJy SMGs with radio-counterparts below the sensitivity of current radio surveys. It is possible that SMGs with no detectable radio counterparts lie at significantly higher redshifts than the radio-identified population ($z\gg 3$, C05; \citealt{younger07}). We cannot test this suggestion with our sample as we have targeted SMGs with known redshifts. However, it has also been suggested that the radio-undetected SMGs lie at similar redshifts to the dominant radio-detected population,  but that they have somewhat lower IR luminosities  and slightly colder characteristic dust temperatures, 
a combination which would result in similar observed 850-$\mu$m fluxes \citep{chapman04}. Following this argument, our radio-detected sample may therefore be biased towards systems with hotter characteristic dust temperatures, either due to more compact star forming regions or the presence of an AGN.  To investigate this possibility, we divide our sample into radio-bright and radio-faint SMG subsets.  We first define a luminosity-complete subsample of SMGs based on their rest-frame radio ($1.4$ GHz) luminosity and use their apparent radio fluxes to identify the radio-bright and radio-faint SMGs for which we can then construct composite spectra. These are shown in Fig.~\ref{Radio_24composites}. 

The radio-faint SMG composite has a narrower $7.7 \mu$m PAH feature that emerges sharply from the continuum, while the radio-bright composite displays a broader, less distinct PAH feature. Considering the line-to-continuum parameter traditionally used to gauge the strength of PAH features ($l/c$; e.g., \citealt{genzel98}), the radio-faint composite displays a higher $7.7 \mu$m--$(l/c)$ relative to the radio-bright composite.  A lower $7.7 \mu$m--$(l/c)$ for the radio-bright composite may suggest a larger obscured AGN contribution. However, the EWs of the $7.7 \mu$m PAH feature in these composites are not significantly different within the measurement uncertainties (see Table\,\ref{compositestab}).  Therefore, overall our observations do not reveal any significant difference in mid-IR properties based on radio-luminosity and we conclude that our radio-detected SMG sample is likely to be representative of the whole SMG population at $z\lesssim 3.5$.

%
%
\begin{figure*}
\plottwo{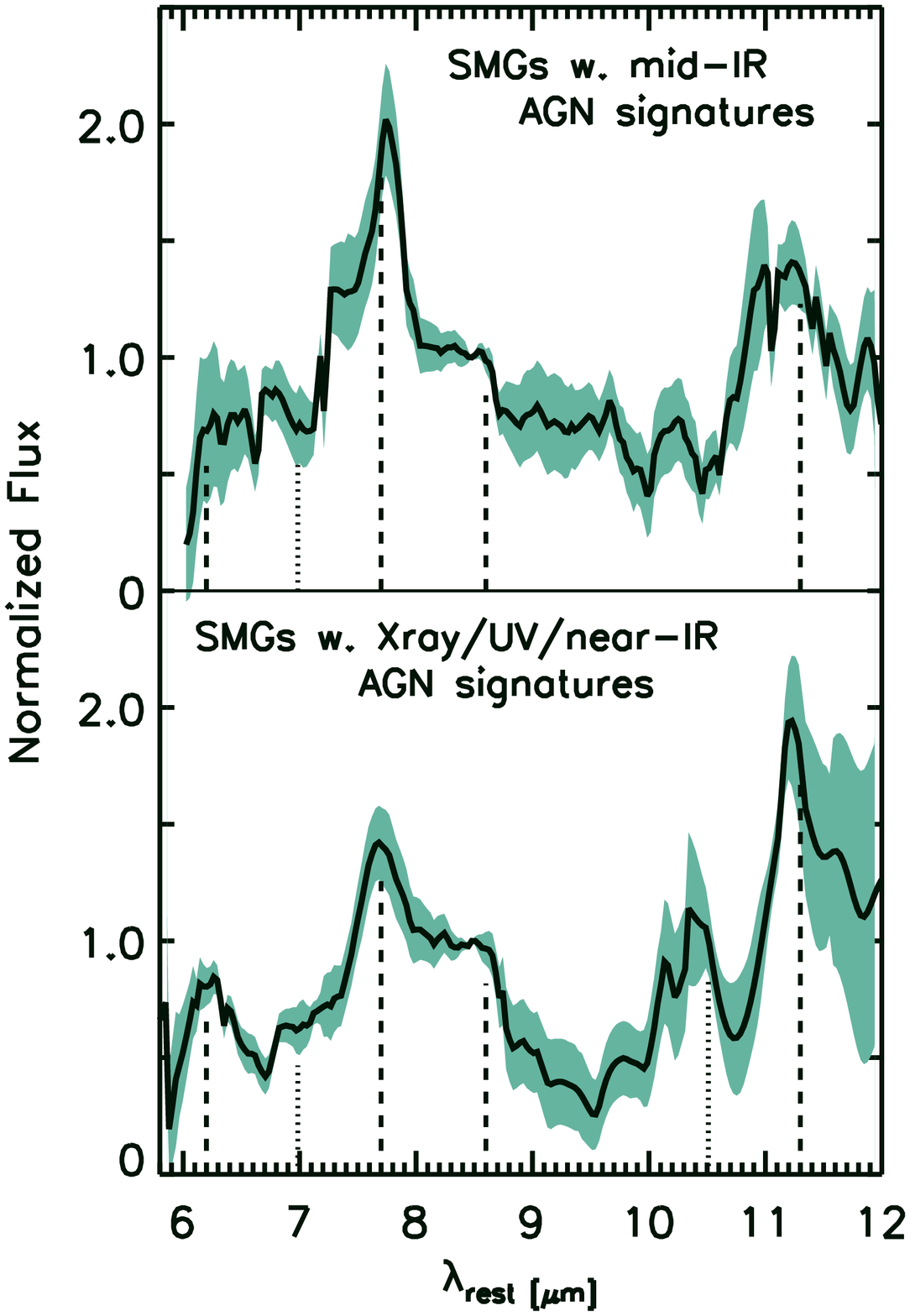}{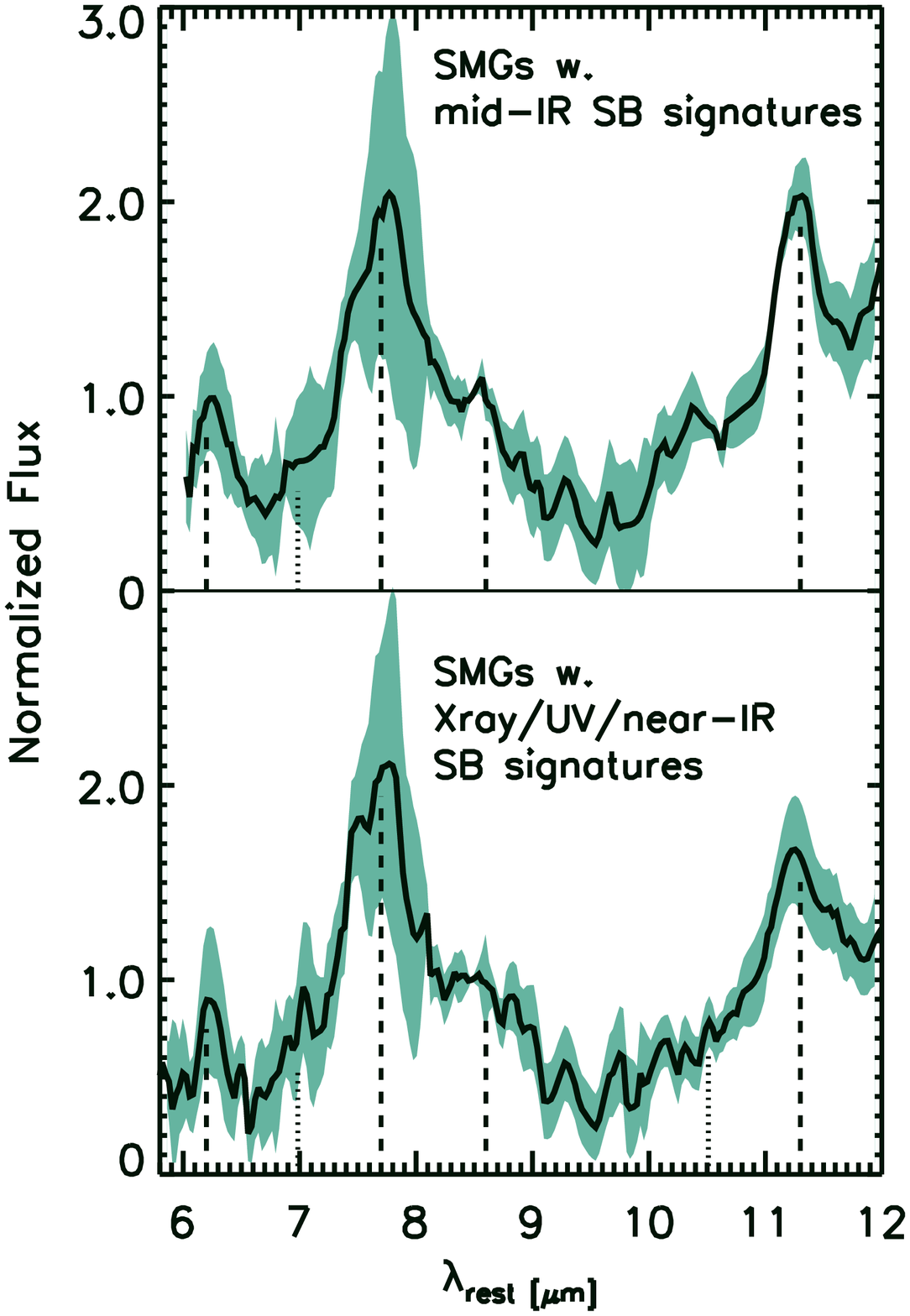}
\caption{({Left:}) The composite spectra of eight SMGs in the sample with AGN signatures either in the X-ray \citep{alexander05}, optical (C05) or near-IR \citep{swinbank04} (bottom) and of the 10 SMGs with mid-IR spectra that suggest the presence of an AGN, either EW$_{7.7\mu m} < 1$ or $\alpha_{MIR} < 0.5$ (top). ({Right:}) Composite spectra of 16 SMGs in our sample with {\it no} AGN signature either in the X-ray, optical or near-IR (bottom) and of the 13 SMGs with mid-IR spectra dominated by strong star formation activity, with EW$_{7.7\mu m} > 1$ and steep mid-IR continuum, $\alpha_{MIR} > 0.5$. We see that the AGN-composites display an enhanced continuum relative to the star-forming counterparts. We also note that the AGN-composite with AGN signatures at wavelengths other than the mid-IR (bottom left panel) displays a prominent emission feature, likely corresponding to [SIV] emission from an AGN. See \S\ref{SBandAGNcomp} for details.
\label{AGN_starburstcomposite}}
\end{figure*}

%
%

\subsubsection{Starburst and AGN-dominated Composites}\label{SBandAGNcomp}

The SMGs in our sample have been classified prior to this work as either AGN- or star-formation--dominated (or starburst) systems  using deep X-ray observations \citep{alexander05}, rest-frame optical spectroscopy  \citep{swinbank04, takata06} and/or rest-frame UV spectroscopy (C05). Table~\ref{signtab} provides a summary of these classifications. We construct composites for an AGN and a star-forming class based on these  classifications to investigate differences in mid-IR spectral properties and plot these in Fig.~\ref{AGN_starburstcomposite}. Since we do not take mid-IR properties into account to build these composites, the differences are not the result of any mid-IR selection bias. 

We find that the composite spectra of both the AGN- and the starburst class display prominent PAH features. However, the PAH emission in the AGN composite lies on top of a stronger continuum, with  a  lower $7.7\mu m$ PAH EW compared with that in the star-forming composite (EW$_{7.7\mu m, AGN}= 1.25 \pm 0.18 \mu$m vs. EW$_{7.7\mu m, SB}=2.09 \pm 0.30 \mu$m). The AGN composite spectrum also has an emission feature at  $\lambda = 10.51 \mu$m, which is much less obvious in the star-forming-composite. This feature may correspond to [S{\sc iv}] narrow-line emission, which is commonly associated with hot starburst activity or the presence of an AGN \citep{spoon02, ogle06}.  We note that all of the individual SMGs which went into this AGN-composite and whose spectra cover the $\lambda \sim 10.5 \mu$m wavelength region show what could correspond to [S{\sc iv}] emission (see Fig.~\ref{spectra}). Since the SMGs that make up this AGN composite have AGN signatures in the rest-frame UV and optical, it is likely that our line of sight has direct access into the broad line region.  Thus it is also possible that this feature corresponds instead to emission from solid silicates very near a central AGN. 

Fig.~\ref{AGN_starburstcomposite} also shows the composite spectra for AGN- and starburst- subsets based on their mid-IR classifications. We classify SMGs with EW$_{7.7 \mu m} \gtrsim 1$ and steep mid-IR continua ($S_{6 \mu m}/S_{12 \mu m} \lesssim 0.3$) as star-formation--dominated systems. SMGs with EW$_{7.7 \mu m} \lesssim 1$ or flat mid-IR continua ($S_{6 \mu m/12 \mu m} \gtrsim 0.3$ or $\alpha_{\rm MIR} \lesssim 0.5$) are classified as `intermediate'. SMGs displaying a continuum-dominated spectrum, either with no sign of PAH features or with a continuum-absorption feature such as SMM\,J163650.43 (see Fig.~\ref{spectra}), we classify as `AGN' to indicate a more significant AGN contribution (see Table~\ref{signtab}). As expected from following these mid-IR classification criteria, the AGN composite in Fig.\,\ref{AGN_starburstcomposite} displays a $7.7\mu$m feature with a lower EW than the starburst composite (EW$_{7.7\mu m, AGN}= 1.57 \pm 0.22 \mu$m vs. EW$_{7.7\mu m, SB}=1.87 \pm 0.26 \mu$m). We observe a similar trend for the other PAH EWs, in particular for the $6.2\mu$m PAH feature, which appears in Fig.\,\ref{AGN_starburstcomposite} to have been heavily diluted by the enhanced continuum in the AGN composite. 

The AGN- and starburst classes, either constructed from mid-IR criteria or from classifications based on observations at other wavelengths, appear to differ mainly in the mid-IR continuum level.  The AGN composites display an enhanced continuum with respect to the starburst composites, resulting in lower $7.7\mu$m PAH EWs. We note that the AGN composites also display prominent and luminous PAH features. Even though the spectra of individual SMGs in our sample display $7.7\mu$m PAH features spanning an order of magnitude in luminosity range, the AGN composites have PAH luminosities within a factor of $\sim 1.5$ of those in the starburst composites. We return to discuss these results in the context of the PAH dilution and PAH destruction scenario in \S\ref{discuss}.

%
%
\begin{figure}
\centering
\includegraphics[scale=0.4, angle=270]{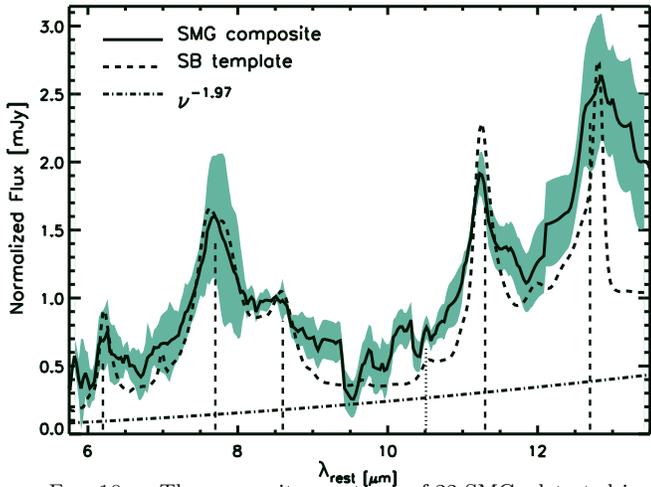}
\caption{The composite spectrum of 22 SMGs detected in our sample as shown in Fig.\,\ref{SMGcomposite}. We fit this composite with the mid-IR composite spectrum of 13 starburst-dominated low-redshift galaxies from \citet{brandl06}, scaled down to match the SMG composite spectrum at 7.7\,$\mu$m. A power-law component with mid-IR spectral slope, $\alpha_{\rm MIR} \sim 1.97$, best represents the continuum in the composite SMG that is unaccounted for by the starburst template. This continuum may arise either from an obscured AGN or optically-thick emission around star-forming regions. See Section\,\ref{agn} for details.
\label{SMGcompositeBrandl}}
\end{figure}

%
%

\subsection{The AGN Contribution to the Bolometric Luminosity in SMGs}\label{agn}

In \citet{md07} we presented the analysis of five SMGs that we also include in this paper. We found that the composite spectrum is well fit by a starburst-like template (represented by M82) with an additional continuum component, likely arising from an AGN.  We found that the mid-IR spectral features of SMGs are more comparable to the starburst M82, than to other well-studied local ULIRGs, such as Arp\,220 and Mrk\,1014. We now consider our complete sample of 24 SMGs and compare the composite SMG spectrum to the composite of nearby nuclear starburst-dominated galaxies (\citealt{brandl06}; see Fig.\,\ref{SMGcompositeBrandl}). By using this starburst template rather than M82, we seek to represent to a better degree the diversity in continuum slopes and PAH strengths in starburst-dominated galaxies in the local Universe.

We find that the mid-IR composite spectrum of the SMGs in our sample is reasonably well fit at $\lambda \lesssim 10 \mu$m by the local starburst template, but we find that an additional power-law continuum is necessary to fully describe the SMG composite spectrum: at the longer wavelengths, $\lambda \gtrsim 10 \mu$m, the starburst template fails to match the continuum level (see Fig.\,\ref{SMGcompositeBrandl}). To quantify this additional continuum that is unaccounted for by the starburst template we fit a power-law to the (SMG composite $-$ starburst-template) residuals. We find that the residuals are best fit by a power-law $F_\nu \sim \nu^{-\alpha_{\rm MIR}}$, where $\alpha_{\rm MIR} = 1.97\pm 0.22$. This additional warm continuum may be associated to obscured AGN activity or to an additional starburst continuum component. We derive an upper limit to the contribution from an AGN to the SMG mid-IR spectrum by making the conservative assumption that this additional continuum arises uniquely from a dust-enshrouded AGN.  

Hard X-rays emission provides one of the most direct routes to estimate the luminosities of AGN.  However, in the presence of high column densities, such as estimated in SMGs (N$_H \gtrsim 10^{23}$\,cm$^{-2}$; \citealt{alexander05}), hard X-ray photons may be completely absorbed.   In these cases an alternate probe for AGN emission is the mid-IR emission from hot dust near the AGN. The mid-IR thus provides a complementary insight into a deeply obscured AGN \citep{sturm06, lutz04}. In particular, \citet{krabbe01} show that the $10 \mu$m-flux is tightly correlated with the X-ray emission, $S_{2-10 \small\textrm{keV}}$, both in Seyfert and in starburst galaxies, with markedly distinct mid-IR slopes. We use the derived correlation for Seyfert galaxies to estimate the X-ray luminosity of the AGN component in the SMG composite spectrum and hence gauge the AGN contribution to the bolometric luminosity of SMGs. 

From the $10 \mu$m continuum flux and the Krabbe et al. relation we estimate an X-ray luminosity, $L_{X} \sim 8 \times 10^{43}$\,erg s$^{-1}$, for the SMGs in our sample at $<\! z\! > \sim 2.0$. Encouragingly, this estimate is in close agreement with the absorption-corrected rest-frame 0.5--8\,keV X-ray luminosities found for the SMGs in the {\it Chandra} Deep Field  \citep{alexander05}, $<\! L_{X}> = 5 \times 10^{43}$ \,erg s$^{-1}$. To assess the contribution of the AGN to the energetics of the SMGs we first estimate the far-IR luminosities of the SMGs
using the observed radio--far-IR correlation for SMGs  \citep{kovacs06}, yielding an average far-IR luminosity of $L_{\rm far-IR} \sim 5 \times 10^{45}$\,erg s$^{-1}$. We then use the approach of \citet{alexander05} and \citet{md07}, and adopt the X-ray-to-FIR luminosity ratio typically found for quasars ($L_X/L_{\rm far-IR} \sim 0.05$; \citealt{alexander05}) to estimate that the AGN in SMGs contribute $\sim 32$\% of the SMG's far-IR luminosities or $L_{\rm far-IR} \sim 1.6 \times 10^{45}$\,erg s$^{-1}$.

We emphasize that this AGN contribution estimate is a strong upper limit for the SMG population. Considering that the enhanced continuum in the mid-IR bright composite suggests that our selection criteria may bias us towards an increased AGN contribution within our sample (see \S\ref{RadioAND24umcomp}), the AGN contribution estimate based on the SMG composite is likely biased high.  It is also very likely that some fraction of the red continuum unaccounted for by the local starburst template arises from dust emission heated by optically-thick starbursts in SMGs, particularly considering the high mid-IR power-law index, $\alpha_{\rm MIR} = 1.97\pm 0.22$, in the range associated with regions dominated by star-formation. Therefore, we quote 32\% as a firm upper limit on the AGN contribution to the bolometric luminosity in typical SMGs.

With an SMG composite dominated by PAH features, it is clear that the bolometric luminosity of SMGs is in general dominated by star-formation.  However, the individual mid-IR SMG spectra display modest variation in AGN contribution and in some cases can reveal the presence of an AGN that remains invisible at other wavelengths (see Table~\ref{signtab}). In particular, there is evidence that SMM\,J123553.26 is a Compton-thick AGN. This SMG is undetected in the X-ray but is bright and dominated by continuum emission in the mid-IR, thus displaying no mid-IR signatures of star formation (see Fig.\,\ref{spectra}).  Extrapolating from its mid-IR spectrum, we estimate a hot dust luminosity\footnote{$\nu$\,L$_{6\mu m} = 4\pi D_L^2 S_{6\mu m} \times \nu_{6\mu m} / (1+z)$}, $L_{6\mu m} \sim2.0 \times 10^{45}$ erg s$^{-1}$. Following the approach described by \citet{alexander08} we use the 0.5--2\,keV observed luminosity of SMM\,J123553.26 \citep{alexander05} to derive a rest-frame 1.6--24.8 keV luminosity and convert it to an intrinsic X-ray luminosity, $L_{2-10 \rm keV} < 2\times10^{42}$ erg s$^{-1}$, assuming an X-ray spectral index, $\Gamma = 1.4$. On the basis of the X-ray--$6\mu$m luminosity of this SMG, we find that it lies significantly below the X-ray--$6\mu$m luminosity relationship found for nearby AGN, suggesting that this source hosts a luminous Compton-thick AGN (e.g., \citealt{alexander08}). However, we note that this SMG lies very close to a second radio source ($\lesssim 5\arcsec$), which complicates the deblending of the individual flux contributions in MIPS $24 \mu$m-imaging \citep{hainline09}. We detect a single continuum trace in the IRS 2D spectrum
that, even though it coincides with the position of SMM\,J123553.26, likely corresponds to the blending of the mid-IR emission of both sources. Therefore, it remains a possibility that the measured mid-IR flux of this source does not correspond to the X-ray source.

We compare the IR luminosity associated with the AGN component in the SMG composite to that of obscured extremely luminous quasars at $z=1.3-3$ studied by \citet{polletta08}. Since the latter have been shown to have $<20\%$ of their bolometric luminosity arising from starburst activity, it is reasonable to assume that the bulk of their IR luminosity arises from AGN activity. We find that the IR emission contributed by the AGN in SMGs is roughly 10\% the IR luminosity of these luminous quasars. Since the bolometric luminosities in these two samples are similar, the respective AGN contributions to the total IR luminosity conform with the evidence that these two populations are dominated by different power sources. Furthermore, this demonstrates that systems dominated by star-formation, such as SMGs, may attain as high luminosities as AGN-dominated systems at high redshift.

%
%

\subsection{Comparison to SMGs in Other Samples}\label{allSMGs}

With a sample of 23 detected SMGs in this work, plus nine from \citep{valiante07} and 13 from \citep{pope08}, we have statistics that allow us to uncover the range in mid-IR properties present within the SMG population.  The majority of SMGs display prominent PAH features atop an underlying weak power-law continuum, potentially arising from an AGN. These spectra suggest starburst-dominated activity, with varying AGN contributions. 

We find only one clear example of a continuum-dominated source with a prominent feature centered at $\lambda \sim 8 \mu$m from unabsorbed continuum, SMM\,J163650.43 \citep{md07}. Three other SMGs in our sample display rather featureless continuum-dominated spectra (SMM\,J105238.30, SMM\,J123553.26, and SMM\,J221737.39). In the sample of nine detected SMGs presented by \citet{valiante07} and \citet{lutz05}, SMM\,J02399-0136 displays PAH features overlaying a particularly strong continuum and is thus classified as a source powered by equal contributions of star formation and a Compton-thick AGN  \citep{valiante07}. Out of the 13 SMGs in the GOODS-North field, SMM\,J123600.15 (C1; \citealt{pope08}) also displays PAH features on a steeply rising continuum and is classified as an SMG with a 44\% AGN contribution to the bolometric luminosity.

We note that 25\% of the SMGs in our sample lie at redshifts $z \lesssim 1.2$, which is a redshift range not covered by the other SMG mid-IR studies.  This extension to lower-redshifts provides us with an insight into the SMG emission at the longer mid-IR wavelengths, $\lambda \gtrsim 11.5 \mu$m.  In Fig.\,\ref{SMGcompositeBrandl} we can see that it is at these longer wavelengths where the local starburst template becomes increasingly insufficient to account for the SMG mid-IR emission and where the additional continuum component becomes increasingly significant. Coverage of this wavelength region within our sample is thus crucial to compare the composite SMG spectrum with potential local analogs and hence to assess the AGN contribution, which we measure to be $<$32\% of the bolometric luminosity. Coverage out to these longer wavelengths is also critical for better constraining the mid-IR continuum slope, giving additional insight to the nature of the underlying power engine based on the measured steepness of the mid-IR spectrum.  Improved continuum constraints allows us in turn to quantify the strength of the 9.7-$\mu$m silicate absorption and investigate the distribution of silicate dust in SMGs, compared to other galaxy populations (see \S\ref{dust}).

%
%
\begin{figure}
\plotone{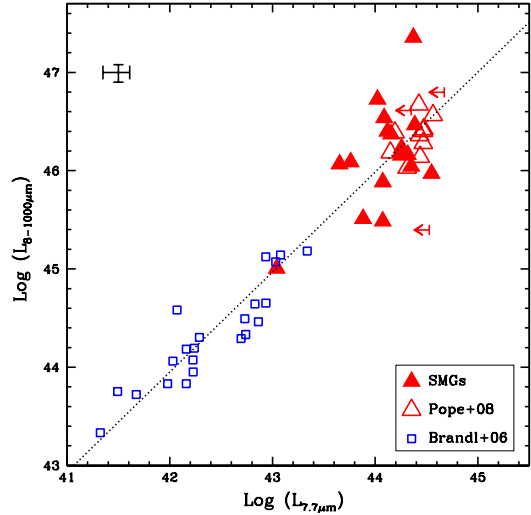}
\caption{PAH luminosities as a function of IR luminosity for the SMGs in our sample and for those in \citet{pope08}. We also include aperture-corrected PAH luminosities of low-redshift nuclear starbursts (\citealt{brandl06}). The line shows the best linear fit to both low-redshift and high-redshift objects (see Eq.\,\ref{eqLIRvsL77}). At low redshifts the  $7.7\mu$m luminosity correlates well with the IR luminosity. At higher redshifts, SMGs appear to mantain such correlation.
\label{SFR}}
\end{figure}

%
%

\section{Discussion}\label{discuss}

We compare the mid-IR properties of the SMGs in our sample with those of starburst-dominated low-redshift galaxies \citep{brandl06}, select local ULIRGs from the Bright IRAS sample \citep{armus06,armus07} and the mid-IR selected sources of \citet{sajina07} and \citet{farrah08}. We use PAH luminosities, PAH relative intensities, the steepness of the continuum and the strength of silicate absorption to compare the level of star-forming activity, the energetics in radiation environments and the distribution of cold and warm dust in these galaxies. 

SMGs have among the most luminous PAH emission in the  universe, displaying levels of activity which are unseen in local samples  (Fig.~\ref{pahVSz}).  This result reflects the significant increase in star-formation activity in the high-redshift SMG population which underpins the thousand-fold increase in the space density of ULIRGs out to $z\sim 2$ (C05). The similarly large PAH luminosities displayed by the mid-IR selected high-redshift objects in \citet{farrah08} also put in evidence the marked increase in PAH luminosities with redshift compared to local ULIRGs.

The presence of large PAH luminosities ($L_{7.7\mu m} \gtrsim 10^{44}$ erg sec$^{-1}$ in SMGs with lower PAH EWs and/or flatter mid-IR continua, such as SMM\,J105238.30 and SMM\,J123549.44, and in SMGs with AGN signatures at other wavelengths, such as SMM\,J123711.98, suggests that even in the presence of an AGN, PAH molecules are abundant in these galaxies. This is also suggested by the large PAH luminosities displayed by the AGN-dominated composite objects discussed in \citet{sajina07} and shown in Fig.~\ref{pahVSz}. It has been suggested, under the PAH {\it destruction} scenario, that highly energetic photons from an AGN overcome the PAH binding energy and reduce the population of PAH molecules and the observed PAH luminosity \citep{voit92}.  However, rather than a decrease in PAH luminosity, we find that SMGs with flatter mid-IR continua and AGN signatures at other wavelengths tend to have lower PAH EWs.   Similar results have been recently discussed in the context of low-redshift ULIRGs by \citet{desai07}, where a decrease in EWs for the 6.2 and $11.3 \mu$m PAH features is observed with increasing $24 \mu$m rest-frame luminosity. These results suggest that an increased AGN contribution results in an enhanced dust continuum that dilutes the prominence of PAH features with respect to the local continuum, rather than in a destruction of PAH molecules. Our study allows us to extend the validity of the PAH-{\it dilution} scenario out to the higher redshifts of SMGs. 

%
%

\subsection{PAH Emission as a Star Formation Rate Indicator}

In the past, authors have explored the use of the mid-IR light as a measure of galaxy SFRs. \citet{elbaz02} claim that the flux measured in the $6.75\mu$m {\it ISO} band is related to the total IR luminosity, which in the absence of an AGN is taken to be a good tracer of star-forming activity \citep{kennicutt98}.  \citet{roussel01} and \citet{forster-schreiber04} showed that there exists a correlation between {\it ISO} $6.75\mu$m light and H$\alpha$ luminosity in local spirals and starburst galaxies. With the advent of Spitzer, $8\mu$m IRAC results from \citet{wu05} and \citet{alonso-herrero06} further confirmed that the $8\mu$m band flux is closely associated with the star-formation taking place in low-redshift ($z \lesssim 0.2$) luminous infrared galaxies with $L_{IR} < 10^{12}$ L$_\odot$.

More recently, several authors have exploited the $6.2 \mu$m PAH emission feature as a SFR indicator, based on the observed correlation between the $6.2 \mu$m PAH and total IR luminosity  (e.g., \citealt{brandl06, pope08}). The wavelength coverage within our study gives us an insight to the $7.7 \mu$m PAH emission feature of 22 SMGs. In Fig.\,\ref{SFR} we show the $7.7 \mu$m PAH luminosities as a function of IR luminosities for SMGs and low-redshift starburst galaxies. We see that $L_{7.7\mu m}$ and $L_{IR}$  for low-redshift starbursts are well correlated and that the SMGs extend this local correlation out to higher redshifts, where we have assumed that the bulk of the total IR luminosity can be associated to star formation (see \S\ref{agn}). We find that the best fit for the complete sample of objects at low- and high-redshifts is:

\begin{equation}\label{eqLIRvsL77}
log(L_{8-1000\mu m}) = (0.99 \pm 0.05) \times log(L_{7.7\mu m}) + (2.45 \pm 2.3).
\end{equation}

The spectral insight to the mid-IR emission of galaxies allows us to disentangle the PAH emission from the AGN contribution to the underlying continuum, which remains otherwise intermixed in mid-IR broad band photometry.  In the presence of an AGN, the continuum contribution may be substantial and even dominate the emission at these mid-IR bands (see for example the continuum-dominated SMGs SMM\,J123553.26, SMM\,J163650.43 and SMM\,J221737.39 in  Fig.~\ref{spectra}). Taking the IR luminosity as a proxy of star-formation activity \citep{kennicutt98}, we use the relation between PAH luminosity and IR luminosity to characterize the $7.7 \mu$m  PAH luminosity as a SFR indicator. Based on this, we find that the $7.7\mu$m PAH luminosity can be used as a SFR indicator following the following form:

\begin{equation}
log(SFR) = -40.9 + 0.99\times log(L_{7.7\mu m}),
\end{equation}

where $L_{7.7\mu m}$  and SFR are in units of erg s$^{-1}$  and M$_\odot$ yr$^{-1}$, respectively. The bulk of SMGs have PAH-derived SFRs $\sim$ {\it few} $\times 10^2-10^3$ M$_\odot$ yr$^{-1}$, consistent with the Xray-derived SFRs by \citet{alexander05} for SMGs in the CDF-North. We obtain a median value of $<SFR>_{7.7\mu m} \sim 800$ M$_\odot$ yr$^{-1}$ for SMGs, including the objects in \citet{pope08}.

%
%

\subsection{Dust Distribution in SMGs}\label{dust}

Our observed silicate optical depths correspond to a median visual extinction of $A_V \sim 6$, assuming $A_V = (18.5 \pm 2) \times \tau_{9.7 \mu m}$ \citep{draine03}. This is significantly higher than the values estimated in previous near-IR studies, $A_V \sim 1-4$, based on the Balmer decrement \citep{takata06} and broad-band optical/near-IR colors \citep{smail04, swinbank04}. This difference demonstrates that these optically-based techniques are only capable of probing the less-obscured regions of the SMGs and hence return lower values for $A_V$. The mid-IR gives us a supplementary insight to the star-formation taking place within the highly optically-obscured regions of SMGs and allows us to quantify the additional extinction at rest-frame optical wavelengths unaccounted for by the Balmer decrement. The difference in values for $A_V$ derived from the mid-IR and the optical/near-IR indicates that H$\alpha$-derived SFRs are (on average) suppressed by a factor of $\gtrsim 100$ when uncorrected for internal extinction. 

The range in silicate optical depths we find in SMGs, $\tau_{9.7\mu m} \sim 0.05$--1.2, falls below the typical values reported for both local ULIRGs and AGNs ($<\! \tau_{9.7 \mu m}\! > \sim 1.5$; \citealt{shi06, hao07}).  Recent work by \citet{desai07} measured the strength of silicate-absorption features in a large sample of low-redshift ULIRGs from the Bright {\it IRAS} Sample. They find that {\it cold} ULIRGs, with steeper mid-IR spectra $S_{25\mu m}/S_{60\mu m} \lesssim 0.2$, have deeper silicate-absorption features than {\it warm} ULIRGs, with $S_{25\mu m}/S_{60\mu m} \gtrsim 0.2$. We find that SMGs have $\tau_{9.7\mu m}$--values close to that of {\it warm} ULIRGs, such as {\it IRAS}\,FSC05189$-$2524, and to the median value found for star-forming low-redshift galaxies ($<\! \tau_{9.7 \mu m}\! > = 0.47 \pm 0.38$; see Fig.~\ref{PAHcontTau}).   

The strength of the silicate feature can provide insight into the distribution of dust along the line of sight. The strong silicate absorption measured for {\it cold} low-redshift ULIRGs can be explained by a small, deeply embedded source with large amounts of obscuring  dust along the line of sight \citep{levenson07}. Shallower silicate-absorption features and the presence of silicate-emission, such as found in the high-redshift mid-IR-selected sources presented by \citet{sajina07} and {\it warm} low-redshift ULIRGs suggest that the mid-IR has a direct view to the hot thermal continuum source, possibly through a clumpy obscuring medium. 

The observed 7.7/6.2 PAH ratios of SMGs (Fig.\,\ref{PAHratios_dist}) are on average lower than those seen in local nuclear starbursts.  This ratio is potentially sensitive to ionization and reddening, with softer radiation fields and lower obscuration leading to lower 7.7/6.2 ratios.  When taken with the differences in silicate absorption between SMGs and local ULIRGs, both these results can be explained by a single cause: lower extinction towards the mid-IR line and continuum emitting regions within SMGs, compared to either local nuclear starbursts or ULIRGs.  The activity in these latter populations are concentrated into highly compact regions and the lower extinction in the SMGs suggests the mid-IR visible star formation in these galaxies is likely to occur in a more extended component, more similar to the activity in the disks of ``normal'' star-forming galaxies in the local Universe. Similar evidence for a more physically extended distribution in starburst-dominated galaxies at $z\sim1.7$ has been reported by \citet{farrah08}.

Evidence for extended star formation within SMGs has also been found at other wavelengths. High-resolution MERLIN/VLA observations of two samples of SMGs have revealed radio morphologies extending out to $\sim 1\arcsec$ or $\sim 8$\,kpc in size \citep{chapman04,biggs08}. These two studies find that $\sim 85$\% of the combined sample of 24 SMGs have extended radio emission, while only $\sim 15$\% appear to be dominated by an unresolved component. Adopting radio emission as a proxy for far-IR emission, these results suggest that massive star-formation is occurring on $\sim 8$\,kpc scales within SMGs. This is in contrast to results for local ULIRGs, where the far-IR emission is confined to a compact nuclear region of $\sim 1$--2\,kpc in size \citep{charmandaris02}. Further evidence for physically extended star formation within SMGs has come from studies of their rest-frame optical emission lines. \citet{swinbank06} mapped H$\alpha$ emission from eight SMGs with Integral Field spectroscopy and found H$\alpha$ structures extending to 1--$2\arcsec$ or $\sim 8$--16\,kpc in size.  These results differ, however, from those of parallel mapping of the molecular gas emission in SMGs. Subarcsecond resolution millimeter imaging of CO emission from six SMGs with the IRAM Plateau de Bure Interferometer suggests compact gas distributions in these galaxies, with sizes $\lesssim 2$\,kpc \citep{tacconi06, tacconi08}. These observations typically target the high-J CO transitions, which trace warm and dense gas, so it remains possible that there are more extended reservoirs of cold gas on $\gg 2$\,kpc scales within these systems which could fuel their extended star formation.  
Our mid-IR results give further evidence towards a more extended distribution of the star-forming regions in SMGs, in contrast to the compact nuclear starbursts seen in similarly luminous local galaxies.  These very extended starbursts may reflect the structural properties of the progenitors of SMGs, which are more gas-rich and may have smaller bulges than those which form local ULIRGs.

%
%

\section{Summary and Conclusions}\label{conc}

SMGs are an enigmatic population which have been proposed to be undergoing tremendous amount of star-forming activity, in order to explain their ultra-luminous infrared emission. The high  stellar masses and SRFs inferred for these galaxies have prompted the idea that SMGs are the likely progenitors of today's most massive galaxies. However, the contribution from AGN activity to the bolometric luminosity has remained a caveat to these results.  We present the largest sample of SMGs observed in the mid-IR with {\it Spitzer} IRS. Our main results are:

1. At the individual level, we find that 80\% of SMGs in our sample display luminous PAH features, with $L_{7.7\mu m} \sim 10^{43}$--10$^{45}$\,erg\,s$^{-1}$. This confirms for a large sample of SMGs that this is a population undergoing intense star-forming activity and that it includes some of the most intense star-formation events ever witnessed in the Universe.

2. We find only four cases of SMGs with continuum-dominated spectra (17\% of our sample), which indicates that even though SMGs appear to be starburst-dominated as an ensemble, some diversity is present within the sample.

3. We quantify the strength of the silicate absorption feature at $\lambda \sim 9.7 \mu$m and, keeping in mind that such measurement requires a careful determination of the continuum, we report lower limits of $\tau_{9.7\mu m}$ only for those SMGs that provide sufficient wavelength coverage for reasonable continuum slope estimates. We find that the distribution in $\tau_{9.7\mu m}$ limits for SMGs falls below that of local ULIRGs and the most obscured low-redshift nuclear starburst-dominated galaxies.  This suggests that SMGs have lower dust obscuration to their mid-IR continuum emitting regions than local ULIRGs. 

4. Comparison of $S_{7.7\mu m}/S_{11.3\mu m}$ PAH flux ratios suggests that SMGs host similar radiation environments to  local star-forming galaxies.  However, the  $S_{7.7\mu m}/S_{6.2\mu m}$ PAH ratio is lower in SMGs than in local nuclear starbursts.  This can most easily be explained if the extinction to these regions is lower in SMGs than in the starburst population.

5. From the composite SMG spectrum we measure the potential AGN contribution as revealed by an enhanced hot-dust continuum at $\lambda_{rest} \gtrsim 10 \mu$m.  We make conservative assumptions regarding the AGN contribution and find that the relative contribution of an AGN with respect to star-formation activity remains low.  With a maximum AGN contribution of $< 32\%$ to the total luminous output, SMGs are clearly a galaxy population dominated by intense starburst activity.  

6. We conclude that the detailed mid-IR spectral properties of SMGs show several differences to local ULIRGs or nuclear starbursts.  These differences can be most easily explained by a difference in the extinction to the mid-IR continuum and line emitting regions of these galaxies, with the SMGs showing systematically lower extinction.  Given the compact geometry of the mid-IR emission from local ULIRGs, the difference in extinction we see argues that the mid-IR emission in SMGs, both continuum and PAH-emission, arises in a more extended component ($> 1$--2\,kpc).  This supports radio and H$\alpha$ studies which indicate that star formation in SMGs is extended on scales of $>1-2$\,kpc, far larger than seen in local ULIRGs.

\acknowledgements

We thank our anonymous referee for useful comments and suggestions. We are very thankful to Alexandra Pope, Elisabetta Valiante, Anna Sajina, Bernard Brandl and Duncan Farrah for facilitating results and/or reduced spectra of their {\it Spitzer} samples for proper comparison to our galaxies in this work. We are also grateful to Laura Hainline, Patrick Ogle, Vandana Desai and James Geach for helpful discussions. KMD is supported by an NSF Astronomy and Astrophysics Postdoctoral Fellowship under award AST-0802399. AWB thanks the Research Corporation and the Alfred P. Sloan Foundation. DMA and IRS acknowledge support from the Royal Society.  This work is based in part on observations made with the {\it Spitzer} Space Telescope, which is operated by the Jet Propulsion Laboratory, California Institute of Technology under a contract with NASA. Support for this work was provided by NASA through an award issued by JPL/Caltech.

\clearpage
\setlength\headsep{0.8in}

%
%
\begin{landscape}

\begin{deluxetable}{lcccccccccccccc}
\tablewidth{0pt}
\tablecolumns{15}
\tabletypesize{\tiny}
\tablecaption{Individual SMGs}

\tablehead{
 \colhead{SMM\,J}  &  \colhead{$z_{C05}$\tablenotemark{a}} & \colhead{$z_{PAH}$\tablenotemark{b}} &  \colhead{$S_{6.2}$\tablenotemark{c}} &  \colhead{$EW_{6.2}$\tablenotemark{d,e}} &  \colhead{$S_{7.7}$} &  \colhead{$EW_{7.7}$} &\colhead{$S_{8.6}$} &  \colhead{$EW_{8.6}$} &\colhead{$S_{11.3}$} &  \colhead{$EW_{11.3}$} &\colhead{$S_{12.7}$} &  \colhead{$EW_{12.7}$} &\colhead{$\tau_{9.7}$\tablenotemark{f}} &  \colhead{$S_{24.0}$\tablenotemark{g}}   \label{resultstab}
}
\startdata
030227.73 & 1.408 & 1.43 & 2.29 $\pm$ 0.46 & 0.43 $\pm$ 0.10 & 8.49 $\pm$ 1.05 & 1.57 $\pm$ 0.25 & 3.29 $\pm$ 0.67 & 0.60 $\pm$ 0.14 & 2.27 $\pm$ 0.47 & 0.41 $\pm$ 0.10 & 2.88 $\pm$ 0.45 & 0.51 $\pm$ 0.10 & $>$ 0.05 & 0.45 \\
030231.81\tablenotemark{h} & 1.316 & -- & $<$1.7 & -- & $<$4.2 & -- & $<$3.5 & -- & $<$0.45 & -- & $<$0.17 & -- & --  & -- \\
105151.69 & 1.147 & 1.62 & 6.13 $\pm$ 0.76 & 1.30 $\pm$ 0.21 & 6.71 $\pm$ 1.17 & 2.08 $\pm$ 0.38 & 2.23 $\pm$ 1.12 & 1.12 $\pm$ 0.43 & 1.68 $\pm$ 0.63 & 1.12 $\pm$ 0.43 & -- & -- & $>$ 0.34 & 0.28 \\
105158.02\tablenotemark{i,j}  & 2.239 & 2.69 & $<$1.06 & $<$0.10 & 8.55 $\pm$ 1.01 & $<$3.85 & 0.69 $\pm$ 0.24 & 0.37 $\pm$ 0.13 & -- & -- & -- & -- & --  & 0.12 \\
105200.22 & 0.689 & 0.67 & -- & -- & -- & -- & -- & -- & 2.10 $\pm$ 0.68 & 0.47 $\pm$ 0.16 & -- & -- & $>$ 0.49 & 0.88 \\
105227.58\tablenotemark{i} & 2.142 & 2.41 & -- & -- & 2.53 $\pm$ 0.51 & 1.96 $\pm$ 0.43 & 0.46 $\pm$ 0.05 & 0.33 $\pm$ 0.20 & -- & -- & -- & -- & --  & 0.19 \\
105238.19 & 1.852 & 1.87 & 3.20 $\pm$ 0.39 & 0.94 $\pm$ 0.15 & 9.15 $\pm$ 0.99 & 3.95 $\pm$ 0.58 & 2.26 $\pm$ 0.48 & 1.24 $\pm$ 0.29 & -- & -- & -- & -- & $>$ 0.27 & 0.51 \\
105238.30\tablenotemark{j}  & 3.036 & 2.99 & $<$4.50 & $<$0.41 & 1.50 $\pm$ 0.21 & 0.14 $\pm$ 0.03 & 0.52 $\pm$ 0.16 & -- & -- & -- & -- & -- & --  & 0.75 \\
123549.44 & 2.203 & 2.24 & -- & -- & 4.80 $\pm$ 0.65 & 0.67 $\pm$ 0.11 & 0.67 $\pm$ 0.07 & 0.10 $\pm$ 0.04 & -- & -- & -- & -- & --  & 0.63 \\
123553.26 & 2.098 & 2.17 & -- & -- & 1.35 $\pm$ 0.44 & 0.27 $\pm$ 0.09 & 1.68 $\pm$ 0.57 & 0.48 $\pm$ 0.17 & -- & -- & -- & -- & --  & 0.39 \\
123707.21 & 2.484 & 2.49 & 1.41 $\pm$ 0.37 & 0.68 $\pm$ 0.19 & 3.53 $\pm$ 0.49 & 1.55 $\pm$ 0.27 & 0.83 $\pm$ 0.22 & 0.35 $\pm$ 0.10 & -- & -- & -- & -- & --  & 0.18 \\
123711.98\tablenotemark{i}  & 1.992 & 1.99 & 2.92 $\pm$ 0.43 & 1.22 $\pm$ 0.22 & 12.06 $\pm$ 1.33 & 7.38 $\pm$ 1.09 & 2.23 $\pm$ 0.49 & 1.73 $\pm$ 0.42 & 1.85 $\pm$ 0.29 & 2.56 $\pm$ 0.47 & -- & -- & $>$ 0.26 & 0.56 \\
123721.87\tablenotemark{i} & 0.979 & 0.97 & -- & -- & 1.40 $\pm$ 0.57 & 0.76 $\pm$ 0.33 & 2.20 $\pm$ 0.55 & 1.41 $\pm$ 0.39 & 2.17 $\pm$ 0.55 & 2.30 $\pm$ 0.64 & 1.40 $\pm$ 0.62 & 1.85 $\pm$ 0.86 & $>$ 1.01 & 0.21 \\
163639.01 & 1.495 & 1.49 & 3.33 $\pm$ 0.52 & 1.27 $\pm$ 0.24 & 4.01 $\pm$ 0.83 & 1.60 $\pm$ 0.35 & 1.79 $\pm$ 0.55 & 0.64 $\pm$ 0.21 & 1.50 $\pm$ 0.62 & 0.51 $\pm$ 0.21 & 2.53 $\pm$ 0.65 & 0.83 $\pm$ 0.22 & $>$ 0.66 & 0.21 \\
163650.43\tablenotemark{k} & 2.378 & -- & -- & -- & $<$10.34 & $<$0.79 & -- & -- & -- & -- & -- & -- & --  & 0.95 \\
163658.78 & 1.190 & 1.20 & -- & -- & 9.37 $\pm$ 1.22 & 1.94 $\pm$ 0.32 & 2.15 $\pm$ 0.68 & 0.44 $\pm$ 0.15 & 2.49 $\pm$ 0.56 & 0.50 $\pm$ 0.13 & 2.26 $\pm$ 0.55 & 0.45 $\pm$ 0.12 & $>$ 0.21 & 0.56 \\
221733.02 & 0.926 & 0.87 & -- & -- & 10.79 $\pm$ 2.48\tablenotemark{l}  & 3.33 $\pm$ 0.63 & 1.27 $\pm$ 0.23 & 0.33 $\pm$ 0.07 & 2.58 $\pm$ 0.26 & 1.35 $\pm$ 0.25 & -- & -- & $>$ 0.11 & 0.31 \\
221733.12 & 0.652 & 0.65 & -- & -- & -- & -- & -- & -- & 8.41 $\pm$ 1.28 & 2.53 $\pm$ 0.46 & 2.89 $\pm$ 0.84 & 0.95 $\pm$ 0.29 & $>$ 1.20 & 0.36 \\
221733.91 & 2.555 & 2.89 & 1.37 $\pm$ 0.34 & 0.62 $\pm$ 0.16 & 2.43 $\pm$ 0.39 & 1.41 $\pm$ 0.27 & -- & -- & -- & -- & -- & -- & --  & 0.24 \\
221735.15\tablenotemark{h,i} & 3.098 & 3.21 & $<$0.77 & -- & 2.82 $\pm$ 0.37 & 1.79 $\pm$ 0.29 & -- & -- & -- & -- & -- & -- & --  & 0.04 \\
221735.84\tablenotemark{i}  & 3.089 & 3.18 & 0.97 $\pm$ 0.18 & -- & 1.53 $\pm$ 0.21 & -- & -- & -- & -- & -- & -- & -- & --  & 0.01 \\
221737.39 & 2.614 & -- & $<$1.07\tablenotemark{h} & $<$0.02 & $<$3.9\tablenotemark{j}  & $<$1.16 & $<$2.54\tablenotemark{j} & $<$0.74 & -- & -- & -- & -- & --  & 0.14 \\
221804.42\tablenotemark{i,j} & 2.517 & 2.55 & 0.77 $\pm$ 0.25 & 0.33 $\pm$ 0.11 & 3.97 $\pm$ 0.72 & 4.52 $\pm$ 0.92 & $<$0.23 & 0.48 $\pm$ 0.63 & -- & -- & -- & -- & --  & 0.10 \\
221806.77 & 3.623 & 1.91 & -- & -- & 8.94 $\pm$ 0.95 & 0.50 $\pm$ 0.07 & 1.69 $\pm$ 0.29 & 0.11 $\pm$ 0.02 & 2.53 $\pm$ 0.34 & 0.35 $\pm$ 0.06 & -- & -- & --  & 0.60 \\
\enddata
\tablenotetext{a}{Optical redshifts from C05}
\tablenotetext{b}{PAH-based Redshifts with a median uncertainty of $\Delta z \sim 0.03$; see \S\ref{redshifts} for details.}
\tablenotetext{c}{PAH fluxes are expressed in units of $10^{-15}$\,erg\,s$^{-1}$cm$^{-2}$. }
\tablenotetext{d}{Quoted errors correspond to {\sc splot} PAH fitting uncertainties and flux uncertainties added in quadrature.}
\tablenotetext{e}{PAH rest-frame EWs are expressed in units of $\mu$m}
\tablenotetext{f}{Optical depth derived from silicate-absorption feature.  See \S\ref{tau} for details.}
\tablenotetext{g}{Fluxes reported are average fluxes within $\lambda_{obs} = 23.7 \pm 2.35 \mu$m  in the individual IRS spectra, to reproduce the $24 \mu$m-MIPS band wavelength coverage.}
\tablenotetext{h}{When PAH features remain undetected above the noise we quote upper limits corresponding to the integrated flux within the PAH fitting windows (see \S\ref{analysis}) in the error spectrum from {\sc spice}.}
\tablenotetext{i}{We detect faint continuum in these SMGs}
\tablenotetext{j}{PAH properties of low-S/N features are quoted as upper limits.}
\tablenotetext{k}{This SMG displays a mid-IR continuum-absorbed bump centered at $\sim 8 \mu$m. We quote the integrated flux within this bump as an upper limit to the $7.7 \mu$m PAH emission.}
\tablenotetext{l}{Only half of the PAH feature falls within the wavelength coverage.}
\end{deluxetable}
\clearpage
\end{landscape}

%
%
\begin{deluxetable*}{lcccccccc}
\tabletypesize{\tiny}
\tablecaption{Composite Spectra}

\tablehead{
 \colhead{Composite}  &   \colhead{\# SMGs} & \colhead{$z_{median}$\tablenotemark{a}} & \colhead{$L_{6.2}$\tablenotemark{b,c}} &   \colhead{$EW_{6.2}$\tablenotemark{c,d}} &  \colhead{$L_{7.7}$} &  \colhead{$EW_{7.7}$} & \colhead{$L_{11.3}$} &  \colhead{$EW_{11.3}$}  
\label{compositestab}}
\startdata
all & 22 & 2.00 & 0.50$\pm$0.06 & 0.15$\pm$0.02 & 3.57$\pm$0.36 & 1.26$\pm$0.18 & 1.16$\pm$0.12 & 0.57$\pm$0.08  \\
24um-Bright\tablenotemark{e} & 7 & 2.10 & 0.76$\pm$0.08 & 0.17$\pm$0.06 & 3.51$\pm$0.35 & 0.99$\pm$0.14 & -- & --  \\
24um-Faint\tablenotemark{f} & 7 & 2.61 & 0.33$\pm$0.03 & $<$0.08 & 10.67$\pm$1.07 & 2.48$\pm$0.35 & -- & --  \\
Radio-Bright\tablenotemark{g} & 5 & 3.04 & -- & -- & 3.67$\pm$0.43 & 0.67$\pm$0.10 & -- & --  \\
Radio-Faint\tablenotemark{h} & 5 & 2.38 & -- & -- & 1.92$\pm$0.22 & 0.58$\pm$0.09 & -- & --  \\
Starburst-dominated (Xray/UV/opt)\tablenotemark{i} & 16 & 1.98 & 0.59$\pm$0.08 & 0.17$\pm$0.03 & 5.46$\pm$0.56 & 2.09$\pm$0.30 & 1.87$\pm$0.24 & 1.31$\pm$0.21  \\
Starburst-dominated (mid-IR)\tablenotemark{j} & 12 & 1.97 & 1.02$\pm$0.12 & 0.27$\pm$0.04 & 5.04$\pm$0.52 & 1.87$\pm$0.26 & 2.09$\pm$0.24 & 1.44$\pm$0.2  \\
AGN (Xray/UV/opt)\tablenotemark{k} & 8 & 2.00 & 0.79$\pm$0.25 & 0.23$\pm$0.04 & 3.23$\pm$0.35 & 1.25$\pm$0.18 & 1.66$\pm$0.19 & 1.11$\pm$0.16  \\
AGN (mid-IR)\tablenotemark{l} & 10 & 2.02 & 0.75$\pm$0.12 & $<$0.19 & 4.62$\pm$0.48 & 1.57$\pm$0.22 & 1.53$\pm$0.18 & 0.92$\pm$0.14  \\
\enddata
\tablenotetext{a}{Median redshift of SMGs comprising composite spectrum.}
\tablenotetext{b}{PAH luminosities are expressed in units of $10^{44}$\,erg\,s$^{-1}$.}
\tablenotetext{c}{Quoted errors correspond to {\sc splot} PAH fitting uncertainties and flux uncertainties added in quadrature.}
\tablenotetext{d}{PAH rest-frame EWs are expressed in units of [$\mu$m].}
\tablenotetext{e}{S$_{24\mu \rm m} > 0.40$\,mJy from the $8 \mu$m luminosity-complete subsample of SMGs with L$_{8 \mu \rm m} > 10^{32}$\,erg\,s$^{-1}$}
\tablenotetext{f}{S$_{24\mu \rm m} < 0.40$\,mJy from the $8 \mu$m luminosity-complete subsample of SMGs with L$_{8 \mu \rm m} > 10^{32}$\,erg\,s$^{-1}$}
\tablenotetext{g}{S$_{1.4 \rm GHz} > 95 \mu$Jy from the luminosity-complete subsample of SMGs with L$_{1.4 \rm GHz} > 2 \times 10^{24}$\,erg\,s$^{-1}$\,Hz$^{-1}$}
\tablenotetext{h}{S$_{1.4 \rm GHz} < 95 \mu$Jy from the luminosity-complete subsample of SMGs with L$_{1.4 \rm GHz} > 2 \times 10^{24}$\,erg\,s$^{-1}$\,Hz$^{-1}$}
\tablenotetext{i}{Composite of SMGs classified as starburst-dominated in the rest-frame optical \citep{swinbank04, takata06}, UV (C05) and/or X-ray \citep{alexander05}}
\tablenotetext{j}{Composite of SMGs classified as starburst-dominated in the mid-IR, with $EW_{7.7\mu \rm m} > 1$ and $\alpha_{\rm MIR} > 0.5$. See text for details.}
\tablenotetext{k}{Composite of SMGs with AGN signatures in the X-ray \citep{alexander05} and/or in the near-IR \citep{swinbank04, takata06}}
\tablenotetext{l}{Composite of SMGs with AGN signatures in the mid-IR: $EW_{7.7\mu \rm m} < 1$ and/or $\alpha_{\rm MIR} < 0.5$. See text for details.}
\end{deluxetable*}
\clearpage

\clearpage
%
%
\begin{deluxetable*}{lrcccccl}
\tabletypesize{\scriptsize}
\tablecaption{AGN and starburst (SB) signatures in Radio-Identified SMGs}
\tablehead{
\colhead{SMM\,J} &  \colhead{$\alpha_{\rm MIR}$} & \colhead{$S_{6.0}/S_{12.0}$\tablenotemark{a}}  & \colhead{X-ray\tablenotemark{b}} & \colhead{UV\tablenotemark{c}} & \colhead{H$\alpha$} & \colhead{mid-IR\tablenotemark{d}} & \colhead{Comment} 
\label{signtab}}
\startdata
030227.73 & 2.08$\pm$0.11& 0.24 $\pm$ 0.12 & -- & SB & AGN\tablenotemark{e,f} & SB & $\alpha_{\rm MIR} > 0.5, EW_{7.7\mu m} > 1$ \\
030231.81 & -- & -- & -- & SB & -- & -- & undetected \\
105151.69 & 0.07 $\pm$ 0.22 & 0.96 $\pm$ 0.48 & -- & SB & -- & int & $\alpha_{\rm MIR} < 0.5, EW_{7.7\mu m} > 1$ \\
105158.02 & -- & -- & -- & SB & SB\tablenotemark{e} & SB? & strong PAH emission above faint continuum \\
105200.22 & 1.76 $\pm$ 0.11 & 0.30 $\pm$ 0.14& -- & SB & -- & SB & $\alpha_{\rm MIR} > 0.5$ \\
105227.58 & -- & -- & -- & SB & -- & SB & $EW_{7.7\mu m} > 1$, faint continuum \\
105238.19 & -- & -- & -- & SB & -- & SB & $EW_{7.7\mu m} > 1$ \\
105238.30 & -- & -- & -- & AGN & -- & AGN & $EW_{7.7\mu m} < 1$, continuum-dominated \\
123549.44 & -- & -- & AGN & SB & int\tablenotemark{e}/AGN\tablenotemark{f} & int & $EW_{7.7\mu m} < 1$ \\
123553.26 & $-$0.75 $\pm$ 0.50 & -- & SB & SB & -- & AGN & $\alpha_{\rm MIR} < 0.5$, $EW_{7.7\mu m} < 1$, continuum-dominated \\
123707.21 & 2.46 $\pm$ 1.3 & -- & AGN & SB & SB\tablenotemark{e} & SB & $EW_{7.7\mu m} > 1$ \\
123711.98 & $-$0.10 $\pm$ 0.30 & $<$1.07 & AGN & SB & -- & int & $\alpha_{\rm MIR}<0.5,EW_{7.7\mu m} > 1$, no continuum \\
123721.87 & 0.22 $\pm$ 0.33 & $<$0.86 & AGN & AGN & SB\tablenotemark{f} & int & $EW_{7.7\mu m} < 1, \alpha_{\rm MIR} < 0.5$,  faint continuum \\
163639.01 & 2.20 $\pm$ 0.32 & $<$0.22 & -- & SB & SB\tablenotemark{e} & SB & $EW_{7.7\mu m} > 1, \alpha_{\rm MIR} > 0.5$ \\
163650.43 & -- & -- & -- & int & AGN\tablenotemark{e,f} & AGN & Absorbed continuum \\
163658.78 & 2.07 $\pm$ 0.32 & 0.24 $\pm$ 0.06 & -- & SB & -- & SB & $\alpha_{\rm MIR} > 0.5, EW_{7.7\mu m} > 1$ \\
221733.02 & $-$0.62 $\pm$ 0.45 & -- & -- & SB & SB\tablenotemark{f} & int & $\alpha_{\rm MIR} < 0.5$, prominent $7.7\mu$m PAH feature \\
221733.12 & 1.26 $\pm$ 0.25 & -- & -- & SB & -- & SB & $\alpha_{\rm MIR} > 0.5$, prominent $11.3, 12.7\mu $m PAH features \\
221733.91 & 0.84 $\pm$ 0.72 & -- & -- & SB & SB\tablenotemark{e} & SB & $\alpha_{\rm MIR} > 0.5, EW_{7.7\mu m} > 1$ \\
221735.15 & -- & -- & -- & SB & -- & SB? & $EW_{7.7\mu m} > 1$, no continuum \\
221735.84 & -- & -- & -- & SB & -- & SB? & $7.7\mu$m PAH feature and no continuum \\
221737.39 & 1.63$\pm$0.54 & -- & -- & SB & AGN\tablenotemark{f} & AGN & featureless, continuum-dominated \\
221804.42 & -- & -- & -- & SB & -- & SB? & $EW_{7.7\mu m} > 1$, no continuum  \\
221806.77 & $-$0.39 $\pm$ 0.35 & -- & -- & SB & -- & int & $EW_{7.7\mu m} < 1$, $\alpha_{\rm MIR} < 0.5$ 
\enddata
\tablenotetext{a}{We report mid-IR colors for SMGs with wavelength coverage $\lambda_{rest} \sim 6$--$12\mu$m.}
\tablenotetext{b}{from \citet{alexander05}}
\tablenotetext{c}{from C05}
\tablenotetext{d}{Classification based on mid-IR spectral properties (see \S\ref{SBandAGNcomp}).}
\tablenotetext{e}{from \citet{swinbank04}}
\tablenotetext{f}{from \citet{takata06}}
\end{deluxetable*}

\end{document}